\documentclass[preprint,12pt,authoryear]{elsarticle}

\usepackage[markup=default]{changes}
\usepackage{booktabs}
\usepackage{xr}
\usepackage{latexsym,bm}
\usepackage{amssymb,amsmath,amsfonts}
\usepackage{graphicx}
\usepackage{url}
\usepackage{algorithm}
\usepackage{algpseudocode}
\usepackage{multirow}
\usepackage{caption}
\usepackage{subcaption}
\usepackage{CJKutf8}

\journal{arXiv}

\begin{document}

\begin{frontmatter}

\title{Reconstructing East Asian Temperatures from 1368 to 1911 Using Historical Documents, Climate Models, and Data Assimilation}

\author[as]{Eric Sun}
\author[ntnu]{Kuan-hui Elaine Lin}
\author[ntu]{Wan-Ling Tseng}
\author[rcec]{Pao K.~Wang}
\author[as]{Hsin-Cheng Huang\corref{cor1}}
\ead{hchuang@stat.sinica.edu.tw}

\address[as]{Institute of Statistical Science, Academia Sinica, Taiwan}
\address[ntnu]{Graduate Institute of Sustainability Management and Environmental Education, National Taiwan Normal University, Taiwan}
\address[ntu]{Ocean Center, National Taiwan University, Taiwan}
\address[rcec]{Research Center for Environmental Changes, Academia Sinica, Taiwan}
\cortext[cor1]{Corresponding author}

\date{}

\begin{abstract}
\baselineskip=15pt
We propose a novel approach for reconstructing annual temperatures in East Asia from 1368 to 1911, leveraging the Reconstructed East Asian Climate Historical Encoded Series (REACHES). The lack of instrumental data during this period poses significant challenges to understanding past climate conditions. REACHES digitizes historical documents from the Ming and Qing dynasties of China, converting qualitative descriptions into a four-level ordinal temperature scale. However, these index-based data are biased toward abnormal or extreme weather phenomena, leading to data gaps that likely correspond to normal conditions. To address this bias and reconstruct historical temperatures at any point within East Asia, including locations without direct historical data, we employ a three-tiered statistical framework. First, we perform kriging to interpolate temperature data across East Asia, adopting a zero-mean assumption to handle missing information. Next, we utilize the Last Millennium Ensemble (LME) reanalysis data and apply quantile mapping to calibrate the kriged REACHES data to Celsius temperature scales. Finally, we introduce a novel Bayesian data assimilation method that integrates the kriged Celsius data with LME simulations to enhance reconstruction accuracy. We model the LME data at each geographic location using a flexible nonstationary autoregressive time series model and employ regularized maximum likelihood estimation with a fused lasso penalty. The resulting dynamic distribution serves as a prior, which is refined via Kalman filtering by incorporating the kriged Celsius REACHES data to yield posterior temperature estimates. This comprehensive integration of historical documentation, contemporary climate models, and advanced statistical methods improves the accuracy of historical temperature reconstructions and provides a crucial resource for future environmental and climate studies.\\
\end{abstract}


\end{frontmatter}

\baselineskip=24pt
\section{Introduction}

Climate change is a complex, multidimensional process with profound implications for the sustainability of our planet. Driven by the interplay between human activities and natural climatic processes, it necessitates comprehensive strategies encompassing mitigation and adaptation. An in-depth understanding of historical climate patterns and their impacts is essential for devising informed and effective strategies to safeguard future generations.

In response to this need, a research team at Academia Sinica has developed an innovative index-based historical climate database: the Reconstructed East Asian Climate Historical Encoded Series (REACHES, \citealp{Wang2018REACHES}). This database utilizes a wealth of Chinese historical documents from the Ming (1368--1644 CE) and Qing (1644--1912 CE) dynasties, offering a unique perspective on paleoclimate in East Asia.

Although the REACHES database offers a wealth of historical context, its geographic and temporal coverage is limited. Because it primarily relies on sites with existing narrative records, data volume is significantly reduced in earlier years and less populated regions. Furthermore, the dataset variables are restricted to four or five ordinal scales. To overcome these challenges, we propose a three-tiered statistical framework to correct these biases and enable a more precise reconstruction of historical temperatures across East Asia.

In the initial phase of our methodology, we employ kriging, the best linear unbiased prediction technique, to interpolate annual temperature data for regions lacking direct records. We adopt a zero-mean assumption based on the hypothesis that areas without data generally experienced normal weather conditions \citep{Wang2024}. Additionally, we treat the index data as interval-censored to address their discontinuous nature. This approach yields a geostatistically smoothed surface across East Asia, providing spatially continuous temperature estimates along with associated uncertainty quantifications.

The second phase involves converting the kriged temperature data into Celsius scales using a quantile mapping technique (\citealt{Cannon2015}). This adjustment aligns the distribution of the kriged data from REACHES with temperature data from the Community Earth System Model Last Millennium Ensemble (LME), a comprehensive global climate model (\citealt{Otto-Bliesner2016}).

Finally, we integrate the calibrated estimates with multiple LME outputs via a Bayesian data assimilation procedure. Data assimilation is widely recognized for its capacity to combine observational data with model simulation outputs to reduce uncertainty and improve estimates \citep{Kalnay2002, Evensen2009}. Various methods exist, such as the Kalman filter \citep{Kalman1960, WANG2021}, variational approaches \citep{Brousseau2011}, and particle filters \citep{Goosse2012}. In our framework, however, a key challenge arises because the prior knowledge is derived from 13 heterogeneous LME time series, exhibiting substantial variability in its mean and variance over time. To accommodate this heterogeneity, we adopt a nonstationary autoregressive model whose parameters are estimated via penalized maximum likelihood with a fused lasso penalty \citep{Tibshirani2004}. The penalty stabilizes abrupt shifts in both the mean and autocorrelation structures, yielding a smoother and more robust prior. Using an efficient coordinate descent algorithm, we optimize these parameters to better capture the varying characteristics of the LME data.

With this refined prior, we perform an offline Kalman filtering and smoothing procedure \citep{Oke2002}, treating the calibrated REACHES data as observational proxies with inherent uncertainties, resulting in posterior reconstructions of temperature that reflect uncertainty contributions from both the data and the LME model.

We validate the final temperature estimates using the earliest instrumental data from the Global Historical Climatology Network (GHCN). The results demonstrate the effectiveness of our Celsius-scaled kriged data and Bayesian data assimilation method in improving temperature predictions.

Our data assimilation approach enhances temperature predictions by integrating REACHES and LME datasets through zero-mean kriging with interval-censored data, quantile mapping, and nonstationary time series modeling. The procedure provides refined estimates of true temperature states, effectively accounting for uncertainties and capturing heterogeneous temporal variations.

The subsequent sections of this article are structured as follows. Section \ref{sec2} introduces the datasets employed in this study. Section \ref{sec3} discusses the application of the best linear predictor on the REACHES data and transforms these data to the Celsius scale. Section \ref{sec4} presents our Bayesian data assimilation approach, and Section \ref{sec5} validates the predicted temperatures. Finally, Section \ref{sec6} offers further discussion.

\section{Data}
\label{sec2}

This study utilizes two primary datasets: the REACHES database derived from historical Chinese documents and the LME data based on numerical climate simulations.

\subsection{REACHES}\label{reachdata}

The REACHES database compiles climate data from Chinese historical documents spanning the Ming and Qing dynasties. It encompasses various climate variables such as precipitation, temperature, thunder, and crop conditions. The database is accessible at \url{https://www.ncdc.noaa.gov/paleo/study/23410}. This study focuses on the temperature data, which are organized chronologically (by year) and geographically (by longitude and latitude). Historical climate databases like REACHES employ ordinal-scale indices to represent qualitative descriptions from historical documents for statistical analysis.

The temperature data in the REACHES dataset are categorized using a four-level ordinal scale that ranges from -2 to 1. This scale is designed to reflect the qualitative assessments in the historical documents, where -2 indicates ``very cold", -1 ``cold", 0 ``normal", and 1 ``warm". This asymmetric scale arises because warm-related records are less frequently documented in Chinese historical texts than cold events, reflecting a cultural emphasis on the more impactful cold weather events in agricultural societies. Table \ref{tab:1} provides examples of how Chinese texts have been converted into these ordinal-scale indices.

\begin{table}[tb]\centering
\begin{tabular}{lllc}
    \toprule
    Year & Chinese Text & English Translation & Temperature Index \\
    \midrule
    1839 & \begin{CJK}{UTF8}{bsmi}
                亢陽，地熱如爐
                \end{CJK}
    & Scorching sun, ground hot as a furnace & 1 \\
    1428 & \begin{CJK}{UTF8}{bsmi}
                秋霜
                \end{CJK}
    & Autumn frost & -1 \\
    1644 & \begin{CJK}{UTF8}{bsmi}
                冬凍，鑿之不入
                \end{CJK}
    & Winter freeze, impenetrable by chisels & -2 \\
    \bottomrule
\end{tabular}
\caption{Examples of historical text translations into ordinal-scale temperature indices in the REACHES dataset.}
\label{tab:1}
\end{table}

\begin{figure}[tb]\centering
    \begin{subfigure}{0.48\linewidth}
        \includegraphics[width=\textwidth]{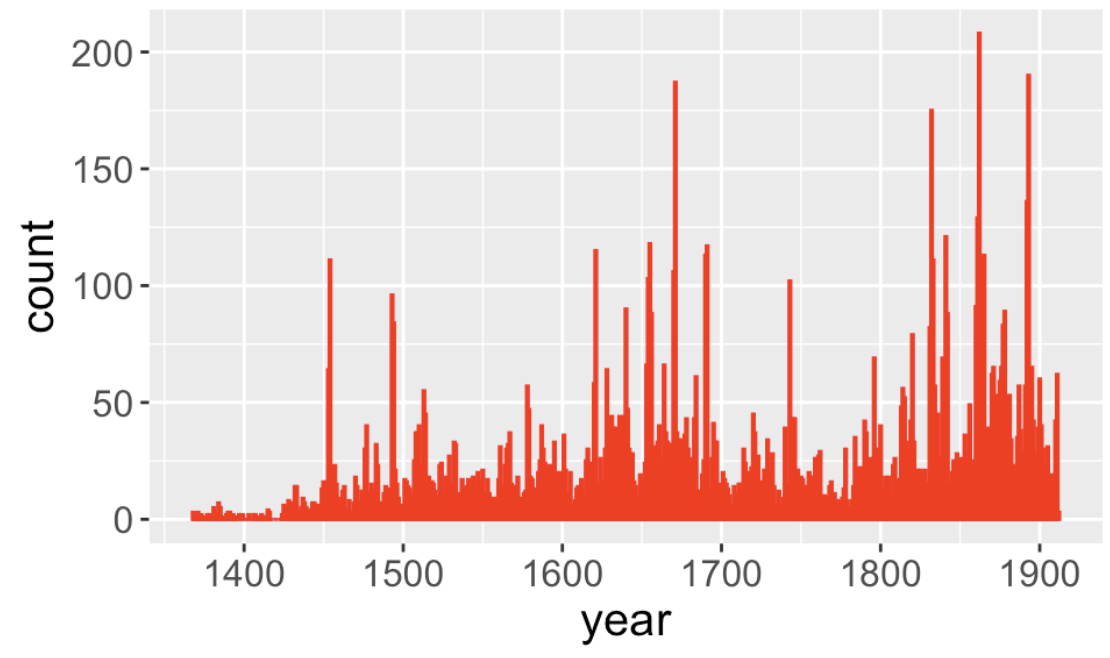}
        \caption{}
    \end{subfigure}
    \hfill
    \begin{subfigure}{0.48\linewidth}
        \includegraphics[width=\textwidth]{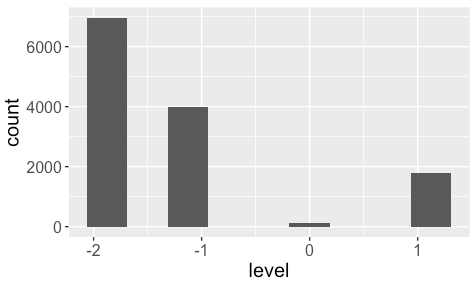}
        \caption{}
    \end{subfigure}
\caption{(a) Annual counts of temperature records in the REACHES dataset from 1368 to 1911; (b) Frequency of each temperature category in the REACHES dataset.}
    \label{ann}
\end{figure}

Figure \ref{ann}(a) displays the time series of annual counts of temperature records in the REACHES dataset, showing a marked increase in data volume in more recent periods compared to earlier years. Historical records tend to emphasize unusual weather events, leading to a skewed distribution of temperature levels, as illustrated in Figure \ref{ann}(b).

\begin{figure}[tb]\centering
    \includegraphics[scale=0.15]{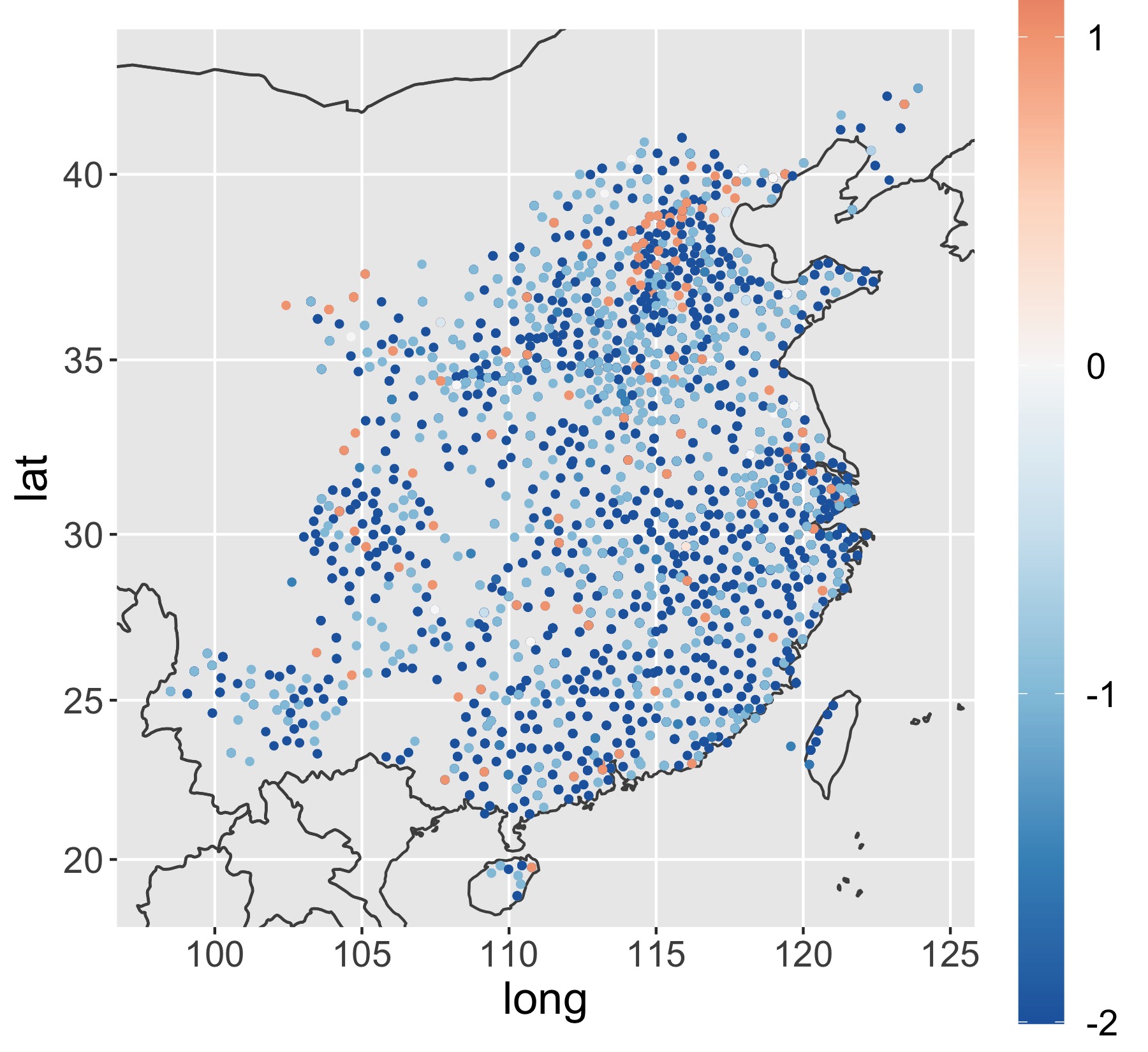}
\caption{Geographical distribution of temperature records from 1368 to 1911 across East Asia, as documented in the REACHES dataset. Each point is color-coded by the average temperature index, showcasing the spatial distribution and density of recorded temperatures.}
\label{eveloc}
\end{figure}

When multiple records from the same location and year are encountered, their temperature indices are averaged to provide a single representative value for that year and location. Figure \ref{eveloc} illustrates the geographical distribution of temperature records across East Asia from the REACHES dataset, spanning from 1368 to 1911. Each point on the map is color-coded according to the average temperature index at that location, effectively highlighting the spatial distribution of records.

\subsection{Last Millennium Ensemble}\label{lmedata}

The LME offers a comprehensive global temperature dataset that extends back to 850 CE. This dataset is indexed by time (year) and geolocation (longitude and latitude) with a spatial resolution of approximately 2 degrees \citep{Otto-Bliesner2016}.
It integrates six time-varying (transient) external forcings, specifically, solar irradiance, greenhouse gas concentrations, aerosols, volcanic emissions, orbital parameters, and land-use changes, into its numerical simulations to recreate Earth’s climate dynamics over the last millennium.

Although the LME relies on sophisticated climate model simulations rather than direct instrumental measurements, potentially raising concerns about the precision of specific annual records, its primary strength lies in its ability to provide reliable estimates over extended periods. The dataset's accuracy is particularly robust over extended timescales, such as decades and centuries, making it highly valuable for historical climate analysis. This characteristic is critical because short-term annual variations may be less precise, but long-term distributions provide reliable insights into climate trends and patterns.

\begin{figure}[ptb]
    \centering
    \begin{subfigure}{0.75\linewidth}
        \includegraphics[width=\textwidth]{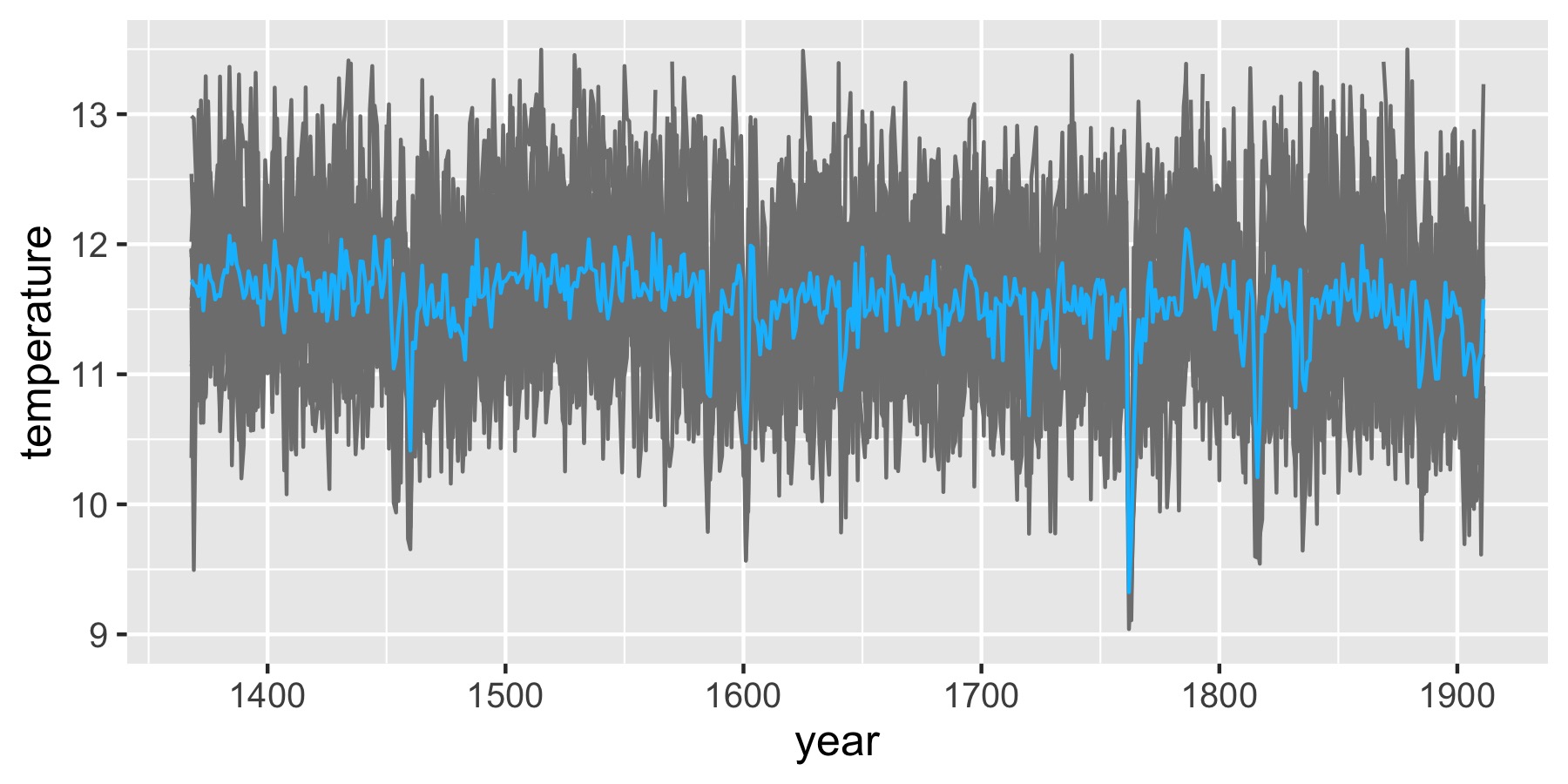}
        \vspace{-0.7cm}
        \caption{Beijing}
        \vspace{0.3cm}
    \end{subfigure}
    \hfill
    \begin{subfigure}{0.75\linewidth}
        \includegraphics[width=\textwidth]{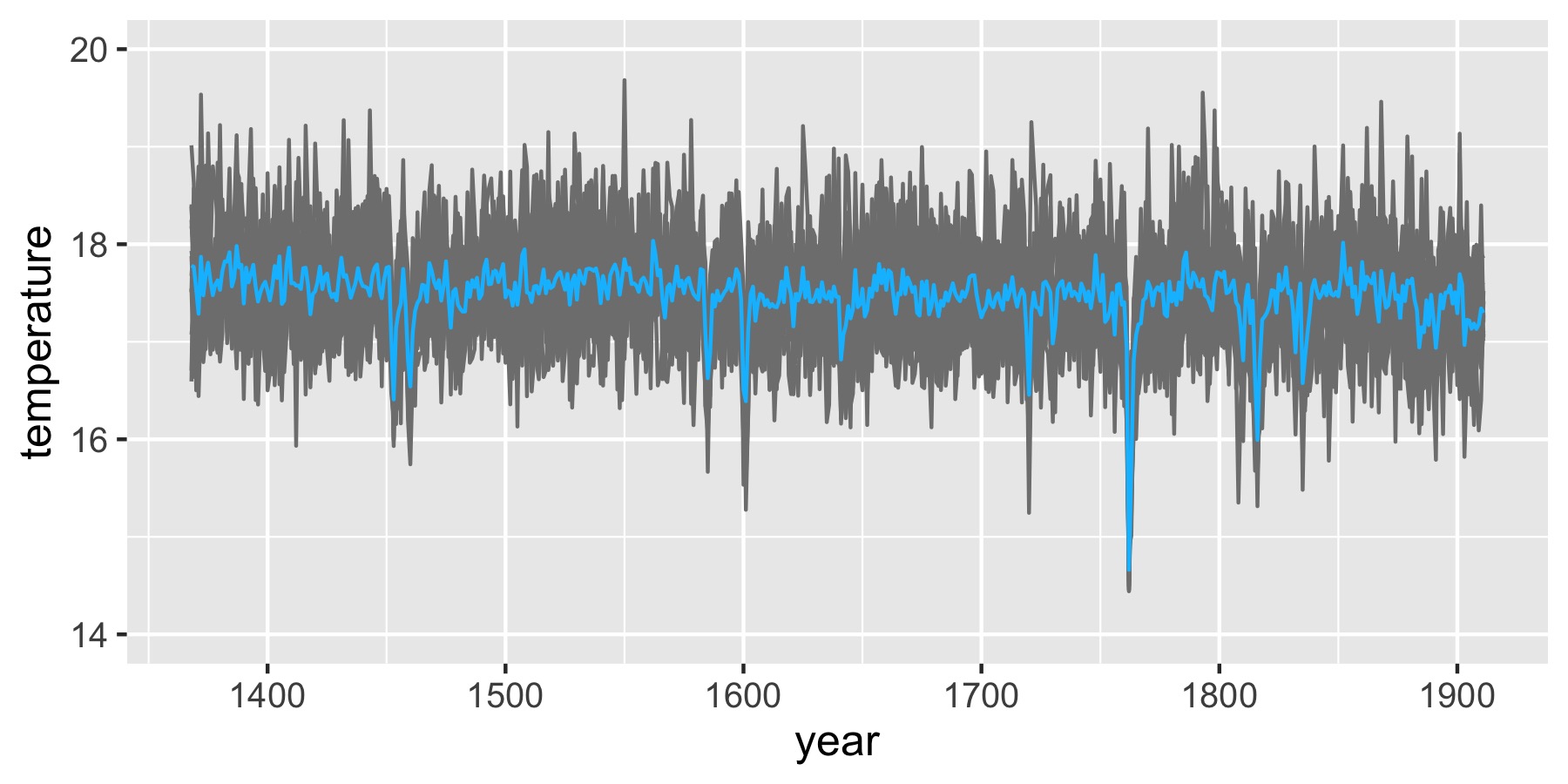}
        \vspace{-0.7cm}
        \caption{Shanghai}
        \vspace{0.3cm}
    \end{subfigure}
    \hfill
    \begin{subfigure}{0.75\linewidth}
        \includegraphics[width=\textwidth]{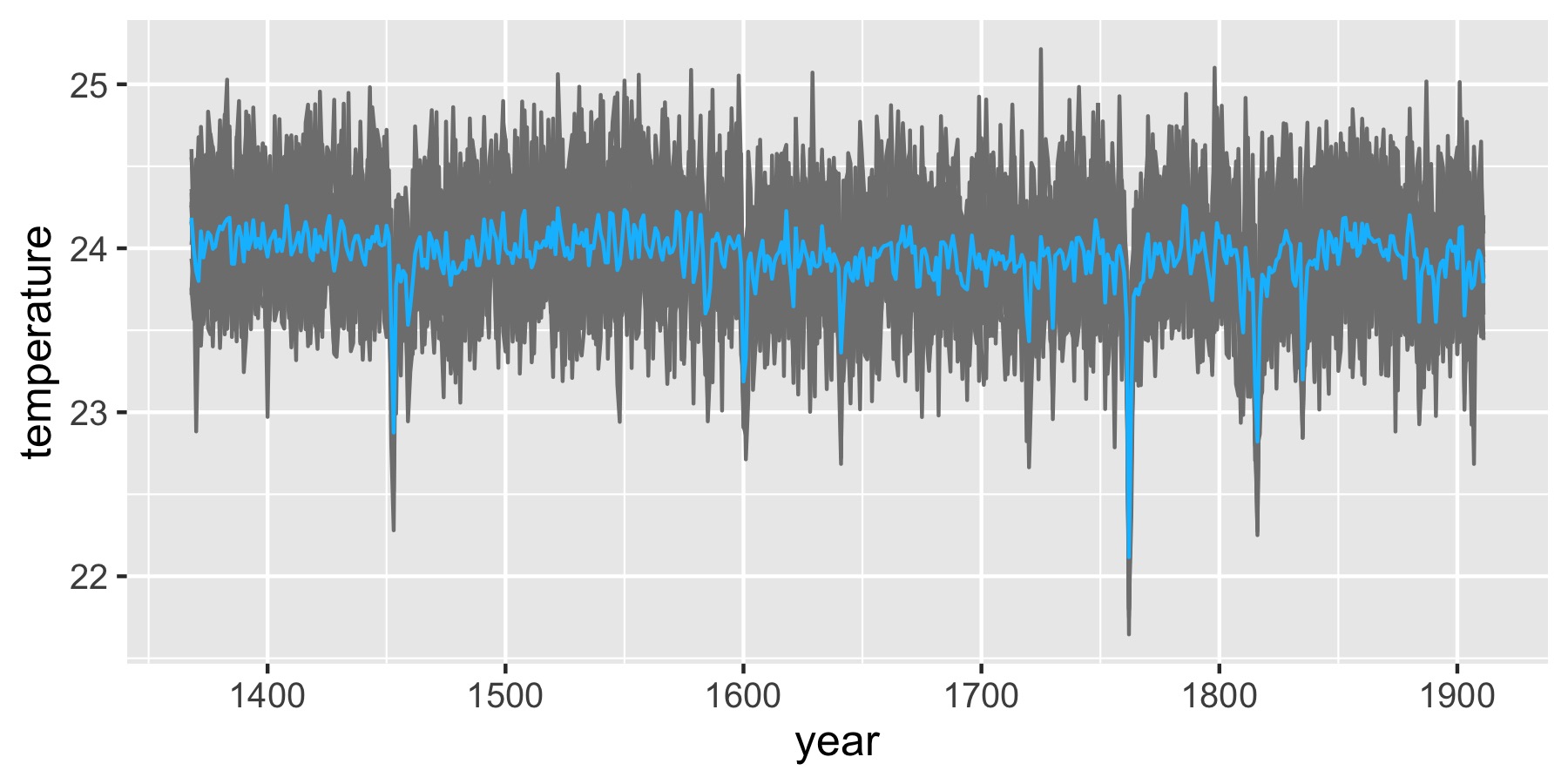}
        \vspace{-0.7cm}
        \caption{Hong Kong}
        \vspace{0.3cm}
    \end{subfigure}
    \caption{Yearly temperature time series data from 13 LME simulations, where the blue curve is the time series averaged over 13 simulations: (a) Beijing; (b) Shanghai; (c) Hong Kong.}
    \label{lmeee}
\end{figure}

In our research, we utilize 13 distinct simulation outputs from the LME, each incorporating all six transient forcings. This captures the full range of natural climate variability and ensures that our prior estimates for historical temperatures reflect diverse potential trajectories. Figure~\ref{lmeee} displays these 13 time series along with their averages for Beijing, Shanghai, and Hong Kong, illustrating the ensemble’s inherent variability. These data serve as a cornerstone in our Bayesian data assimilation framework, providing a prior for subsequent statistical refinement. The LME dataset is publicly available at \url{https://www.cesm.ucar.edu/community-projects/lme}.

\section{Historical Temperature Reconstruction based on REACHES Data}
\label{sec3}

\subsection{Kriging with Zero-Mean Spatial Processes based on Interval-Censored Data}

The REACHES dataset, derived from historical documents, contains numerous gaps in data for certain years or locations, primarily due to its emphasis on exceptional climatic events. This characteristic likely introduces biases into the dataset, emphasizing extremes over typical weather patterns.

To effectively model these data, we employ a zero-mean spatial Gaussian process $\{Y_t(\boldsymbol{s}): \boldsymbol{s} \in D\}$, representing the underlying continuous temperature index for the year $t$ over a geographical region $D \subset \mathbb{R}^2$. Opting for a zero-mean model reflects the assumption that normal weather conditions, which are more frequent but less often documented, constitute the baseline state. Given the annual nature of the data, we consider the temporal dependence among $Y_t(\cdot)$ for years 1368 to 1911 to be minimal and hence treat these processes as independent.  Therefore, for notational simplicity, we drop the subscript $t$ and consider the zero-mean Gaussian spatial process $\{Y(\boldsymbol{s}): \boldsymbol{s} \in D\}$. We assume that the process $Y(\cdot)$ is stationary and has the following exponential covariance function:
\begin{equation}
C_{\mathrm{Y}}(\boldsymbol{s}, \boldsymbol{s}^*) \equiv \mathrm{cov}(Y(\boldsymbol{s}), Y(\boldsymbol{s}^*)) = \sigma_Y^2 \exp \biggl(-\frac{\|\boldsymbol{s} - \boldsymbol{s}^*\|}{\alpha} \biggl);\quad\boldsymbol{s},\boldsymbol{s}^{*} \in D,
\label{eq:exp}
\end{equation}

\noindent where $\sigma_Y^2=\mathrm{var}(Y(\boldsymbol{s}))$ and $\alpha$ is a scale parameter. 

The REACHES temperature data, denoted by $\boldsymbol{Z} \equiv (Z(\boldsymbol{s}_1),\dots, Z(\boldsymbol{s}_{n}))'$, are collected at locations $\boldsymbol{s}_1, \dots, \boldsymbol{s}_n \in D$. We assume that they are observed by rounding the latent variables:
\[
Z^*(\boldsymbol{s}_i)\equiv Y(\boldsymbol{s}_i) + \epsilon(\boldsymbol{s}_i);\quad i=1,\dots,n,
\]
to the nearest integer values in $\{-2, -1, 0, 1\}$:
\begin{equation}     
Z(\boldsymbol{s}_i) = h(Z^*(\boldsymbol{s}_i));\quad i=1,\dots,n,
\label{eq:data}
\end{equation}

\noindent where $\epsilon(\boldsymbol{s}_i) \sim N(0, \sigma^2_\epsilon)$; $i=1,\dots,n$, are white noise, representing measurement errors, and
\[
h(x) = \left\{
\begin{array}{ll}
-2; & \mbox{if }x<-3/2,\\
-1; & \mbox{if }x\in[-3/2,-1/2),\\
0; & \mbox{if }x\in[-1/2,1/2],\\
1; & \mbox{if }x>1/2,\\
\end{array}
\right.
\]
is the rounding function.

Given any location $\boldsymbol{s}_0\in D$, we apply the best linear predictor to estimate $Y(\boldsymbol{s}_0)$:
\begin{equation}
\hat{Y}(\boldsymbol{s}_0) =  \boldsymbol{c}_{{YZ}}(\boldsymbol{s}_0)\boldsymbol{\Sigma}_Z^{-1} (\boldsymbol{Z} - \mu_Z\boldsymbol{1}),
\label{eq:kriging}
\end{equation}

\noindent where $\mu_Z = \mathrm{E}(Z(\boldsymbol{s}_i))$, $\boldsymbol{1}\equiv(1,\dots,1)'$,  $\boldsymbol{c}_{{YZ}}(\boldsymbol{s}_0)\equiv(\mathrm{cov}(Y(\boldsymbol{s}_0),Z(\boldsymbol{s}_1)), \dots, \mathrm{cov}(Y(\boldsymbol{s}_0),Z(\boldsymbol{s}_n)))'$, and $\boldsymbol{\Sigma}_Z$ is an $n \times n$ matrix with the (i, j)-th element $\mathrm{cov}(Z(\boldsymbol{s}_i),Z(\boldsymbol{s}_j))$, for $i,j = 1, \dots, n$ \citep[see][]{Cressie1993}. The corresponding mean-squared prediction error (MSPE) is given by
\begin{equation}
\mathrm{E}\big(\hat{Y}(\boldsymbol{s}_0)-Y(\boldsymbol{s}_0)\big)^2=C_{\mathrm{Y}} (\boldsymbol{s}_0, \boldsymbol{s}_0) - \boldsymbol{c}_{{YZ}}(\boldsymbol{s}_0)' \boldsymbol{\Sigma}_Z^{-1} \boldsymbol{c}_{{YZ}}(\boldsymbol{s}_0).
\label{eq:MSPE}
\end{equation}

Due to the rounding of the latent variables \(Z^*(\boldsymbol{s}_i)\) into discrete indices, the terms \(\mu_Z\), \(\boldsymbol{\Sigma}_Z\), and \(\boldsymbol{c}_{YZ}\) in \eqref{eq:kriging} and \eqref{eq:MSPE} become complex functions of the model parameters \(\alpha\), \(\sigma_Y^2\), and \(\sigma_\varepsilon^2\). Since $Z^*(\boldsymbol{s}_i) \sim N(0, \sigma_Y^2 + \sigma^2_{\epsilon})$, the cumulative distribution function (CDF) of \(Z^*(\boldsymbol{s}_i)\) is then given by
\[
F^*(x)\equiv P(Z^*(\boldsymbol{s}_i)\leq x)=\Phi\big(x\big/(\sigma_Y^2+\sigma_\varepsilon^2)^{1/2}\big),
\]
where \(\Phi(\cdot)\) denotes the standard normal CDF. Thus
\begin{align*}
  \mu_Z
=&~ -2 F^*(-3/2) - (F^*(-1/2) - F^*(-3/2)) + (1 - F^*(1/2)) \\
=&~ -F^*(-3/2) = - \Phi\big(-3\big/\big\{2(\sigma_Y^2+\sigma_\varepsilon^2)^{1/2}\big\}\big).
\end{align*}

The covariance terms \(\boldsymbol{c}_{YZ}(\alpha, \sigma_Y^2, \sigma_\varepsilon^2)\) and \(\boldsymbol{\Sigma}_Z(\alpha, \sigma_Y^2, \sigma_\varepsilon^2)\) in equations \eqref{eq:kriging} and \eqref{eq:MSPE} lack closed-form expressions due to interval censoring but can be numerically evaluated using properties of interval-censored normal distributions. Specifically, the element \(\mathrm{cov}\big( Y(\boldsymbol{s}_0), Z(\boldsymbol{s}_i) \big)\) involves integration over the intervals defined by \(Z(\boldsymbol{s}_i)\). Similarly, for \(\boldsymbol{\Sigma}_Z\), each element \(\mathrm{cov}\big( Z(\boldsymbol{s}_i), Z(\boldsymbol{s}_j) \big)\) requires computing the covariance between two interval-censored variables, necessitating evaluations over two-dimensional regions defined by the observed indices. These numerical evaluations, standard in the analysis of interval-censored data, can be efficiently performed using statistical software that supports algorithms for multivariate normal probabilities and expectations  \citep{Genz2009}.

By treating the rounding of \(Z^*(\boldsymbol{s}_i)\) as interval censoring, we appropriately account for the uncertainty introduced by discretization. This approach allows us to accurately estimate \(\mu_Z\), \(\boldsymbol{c}_{YZ}\), and \(\boldsymbol{\Sigma}_Z\), enabling the application of kriging methods to the REACHES dataset.

Due to interval censoring, we estimate the model parameters \(\alpha\), \(\sigma_Y^2\), and \(\sigma_\varepsilon^2\) in \eqref{eq:exp} and \eqref{eq:data} using a two-step procedure. In the first step, we compute the empirical variograms by pooling all the data from all years and fit the exponential variogram model with a nugget effect using weighted least squares \citep{Cressie1985}. This provides initial estimates \(\tilde{\alpha}\), \(\tilde{\sigma}_Y^2\), and \(\tilde{\sigma}_\varepsilon^2\), obtained without considering interval censoring.

In the second step, we account for interval censoring by deriving bias-corrected functions:
\begin{align}
  f_1(\sigma_Y^2+\sigma_\varepsilon^2)
\equiv&~ \mathrm{var}(Z(\boldsymbol{s})),
\label{eq:f1}\\
  f_2(\sigma_\varepsilon^2)
\equiv&~ \displaystyle\lim_{\|\boldsymbol{s}-\boldsymbol{s}'\|\rightarrow 0}\mathrm{var}(Z(\boldsymbol{s})-Z(\boldsymbol{s}'))/2,
\label{eq:f2}\\
  f_3(\alpha)
\equiv&~ \mathop{\arg\min}_{a>0}\int_0^\infty\big\{C_Z(x;a)- \sigma_Y^2\exp(-x/\alpha)\big\}^2 dx,
\label{eq:f3}
\end{align}

\noindent where
\[
C_Z(x; \alpha) \equiv \mathrm{cov}\big( h(Z^*(\boldsymbol{s})), h(Z^*(\boldsymbol{s}')) \big), \quad \alpha > 0, \, x = \|\boldsymbol{s} - \boldsymbol{s}'\| > 0.
\]
Since the right-hand sides of \(f_1(\cdot)\), \(f_2(\cdot)\), and \(f_3(\cdot)\) in \eqref{eq:f1}--\eqref{eq:f3} can be estimated by \(\tilde{\sigma}_Y^2 + \tilde{\sigma}_\varepsilon^2\), \(\tilde{\sigma}_\varepsilon^2\), and \(\tilde{\alpha}\), respectively, we obtain our final estimates of \(\alpha\), \(\sigma_Y^2\), and \(\sigma_\varepsilon^2\) as follows:
\[
\hat{\sigma}_Y^2 = f_1^{-1}(\tilde{\sigma}_Y^2 + \tilde{\sigma}_\varepsilon^2) - f_2^{-1}(\tilde{\sigma}_\varepsilon^2), \quad \hat{\sigma}_\varepsilon^2 = f_2^{-1}(\tilde{\sigma}_\varepsilon^2), \quad \hat{\alpha} = f_3^{-1}(\tilde{\alpha}).
\]

We proceed to derive \(f_1(\cdot)\), \(f_2(\cdot)\), and \(f_3(\cdot)\), with details provided below.
\begin{itemize}
    \item Derivation of \(f_1(\sigma_Y^2 + \sigma_\varepsilon^2)\): For given values of \(\sigma_Y^2\) and \(\sigma_\varepsilon^2\), we use a Monte Carlo (MC) method by generating an independent and identically distributed (i.i.d.) MC sample of \(Z(\boldsymbol{s})\). This allows us to estimate \(\mathrm{var}(Z(\boldsymbol{s}))\) under interval censoring.
    \item Derivation of \(f_2(\sigma_\varepsilon^2)\): With a given value of \(\sigma_\varepsilon^2\), we compute \(\mathrm{var}\big( h(Y + \epsilon_1) - h(Y + \epsilon_2) \big) / 2\) using an MC sample, where \(Y \sim N\big(0, f_1^{-1}(\tilde{\sigma}_Y^2 + \tilde{\sigma}_\varepsilon^2) - \sigma_\varepsilon^2\big)\) and \((\epsilon_1, \epsilon_2)' \sim N\big( \boldsymbol{0}, \sigma_\varepsilon^2 \boldsymbol{I} \big)\) independently. This step accounts for the measurement error variance in the nugget effect.
    \item Derivation of \(f_3(\alpha)\): For a given value of \(\alpha\), we generate an MC sample of \(Z^*(\boldsymbol{s})\) with \(\sigma_Y^2\) and \(\sigma_\varepsilon^2\) given by \(\hat{\sigma}_Y^2\) and \(\hat{\sigma}_\varepsilon^2\), respectively. We then compute \(C_Z(x; \alpha)\) and minimize the integral in \eqref{eq:f3} to solve for \(\alpha\).
\end{itemize}

\begin{figure}[tb]\centering
    \begin{subfigure}{0.34\linewidth}
        \includegraphics[width=\textwidth,trim={0cm 0cm 0cm 0cm},clip]{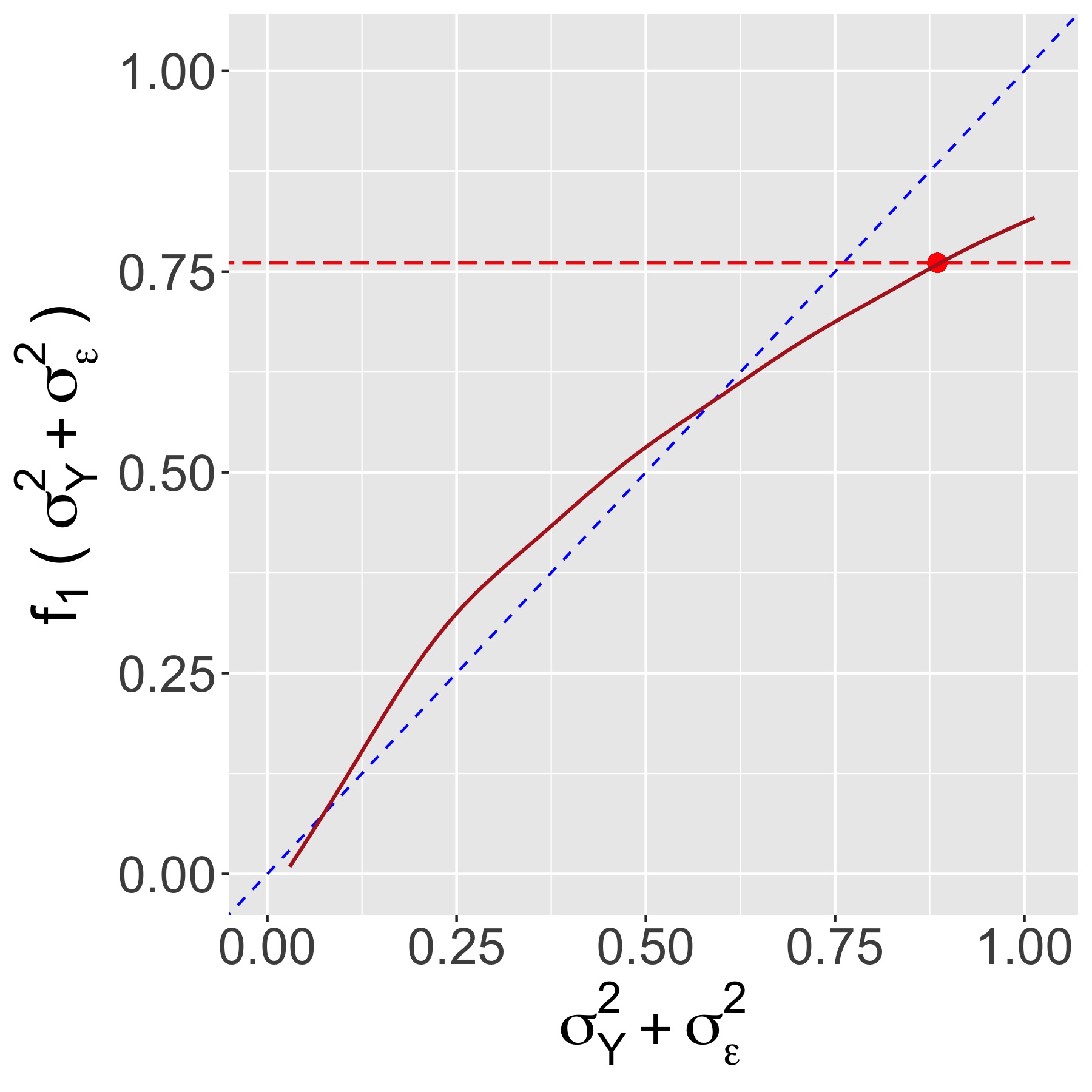}
        \caption{}
    \end{subfigure}
    \begin{subfigure}{0.34\linewidth}
        \includegraphics[width=\textwidth,trim={0cm 0cm 0cm 0cm},clip]{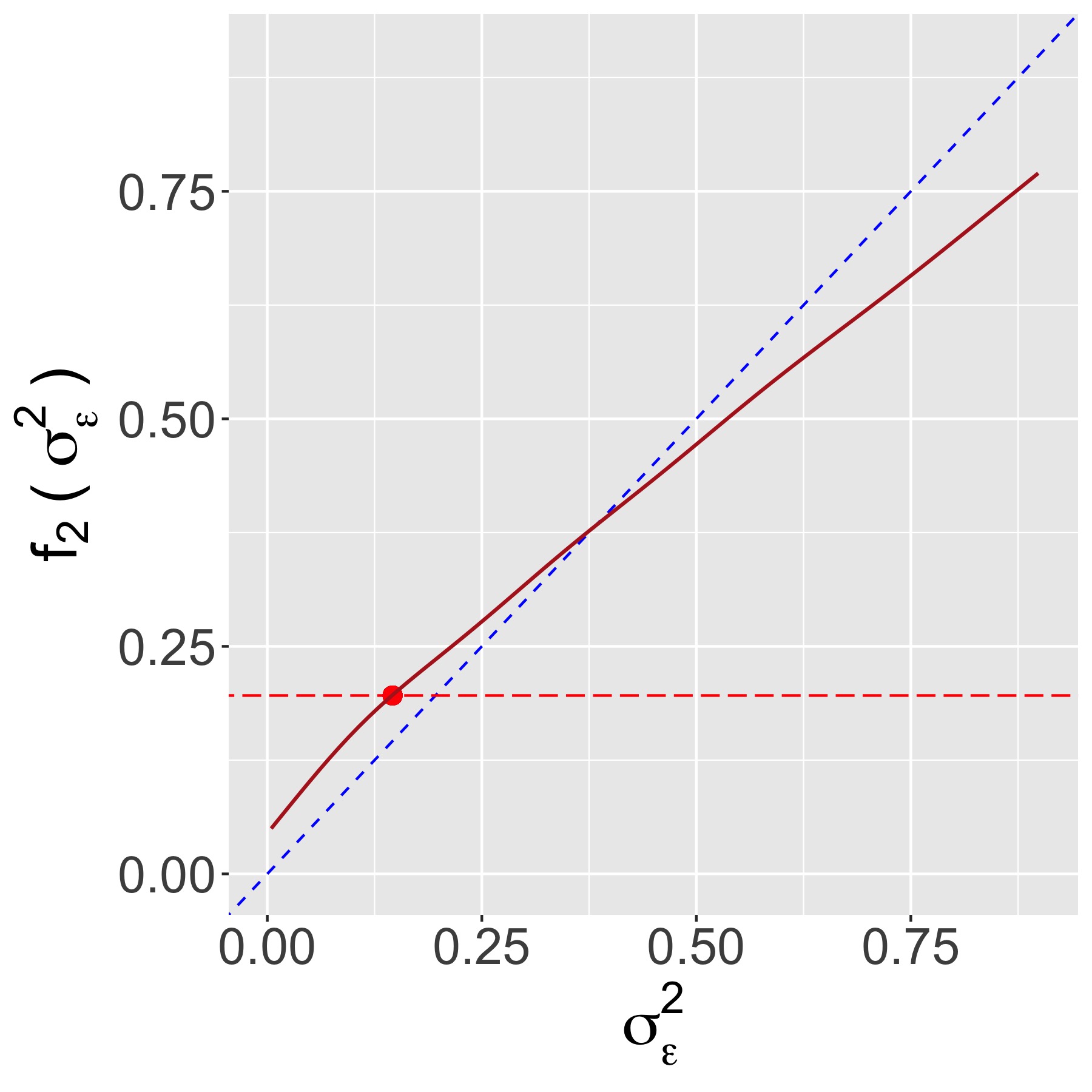}
        \caption{}
    \end{subfigure}
    \begin{subfigure}{0.29\linewidth}
        \includegraphics[width=\textwidth,trim={0cm 0cm 0cm 1.5cm},clip]{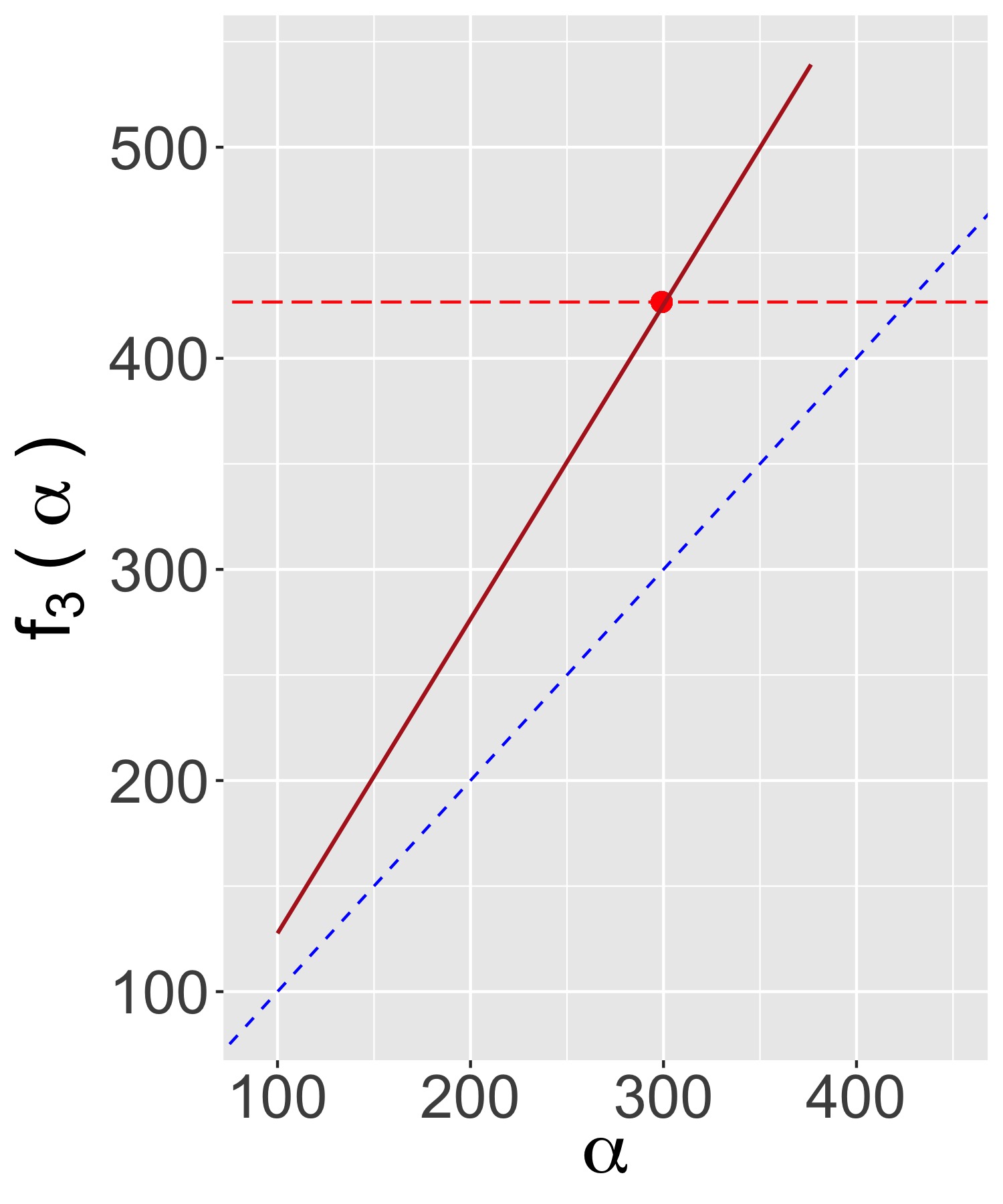}
        \caption{}
    \end{subfigure}
    \caption{Calibration functions for model parameters: (a) \(f_1(\cdot)\) for \(\sigma_Y^2+\sigma_\varepsilon^2\), (b) \(f_2(\cdot)\) for \(\sigma_\varepsilon^2\), and (c) \(f_3(\cdot)\) for \(\alpha\), each represented by a red solid line. Red horizontal dashed lines indicate the initial estimates for each parameter. The blue dashed lines at 45 degrees serve as reference lines to assess deviations from true values.} 
    \label{Monte carlo}
\end{figure}

\begin{figure}[tb]\centering
    \begin{subfigure}{0.49\linewidth}
        \includegraphics[width=\textwidth]{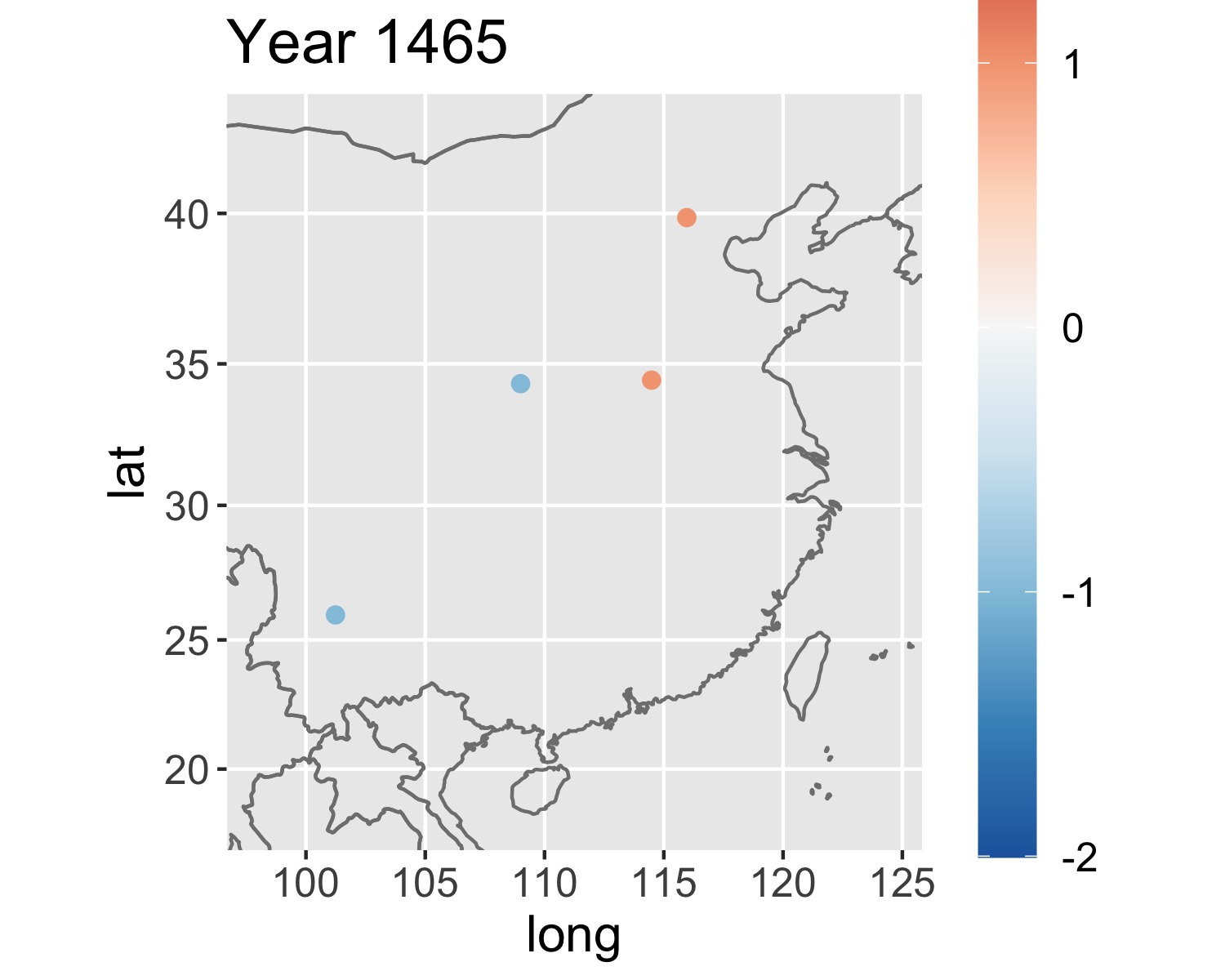}
        \caption{}
    \end{subfigure}
    \hfill
    \begin{subfigure}{0.49\linewidth}
        \includegraphics[width=\textwidth]{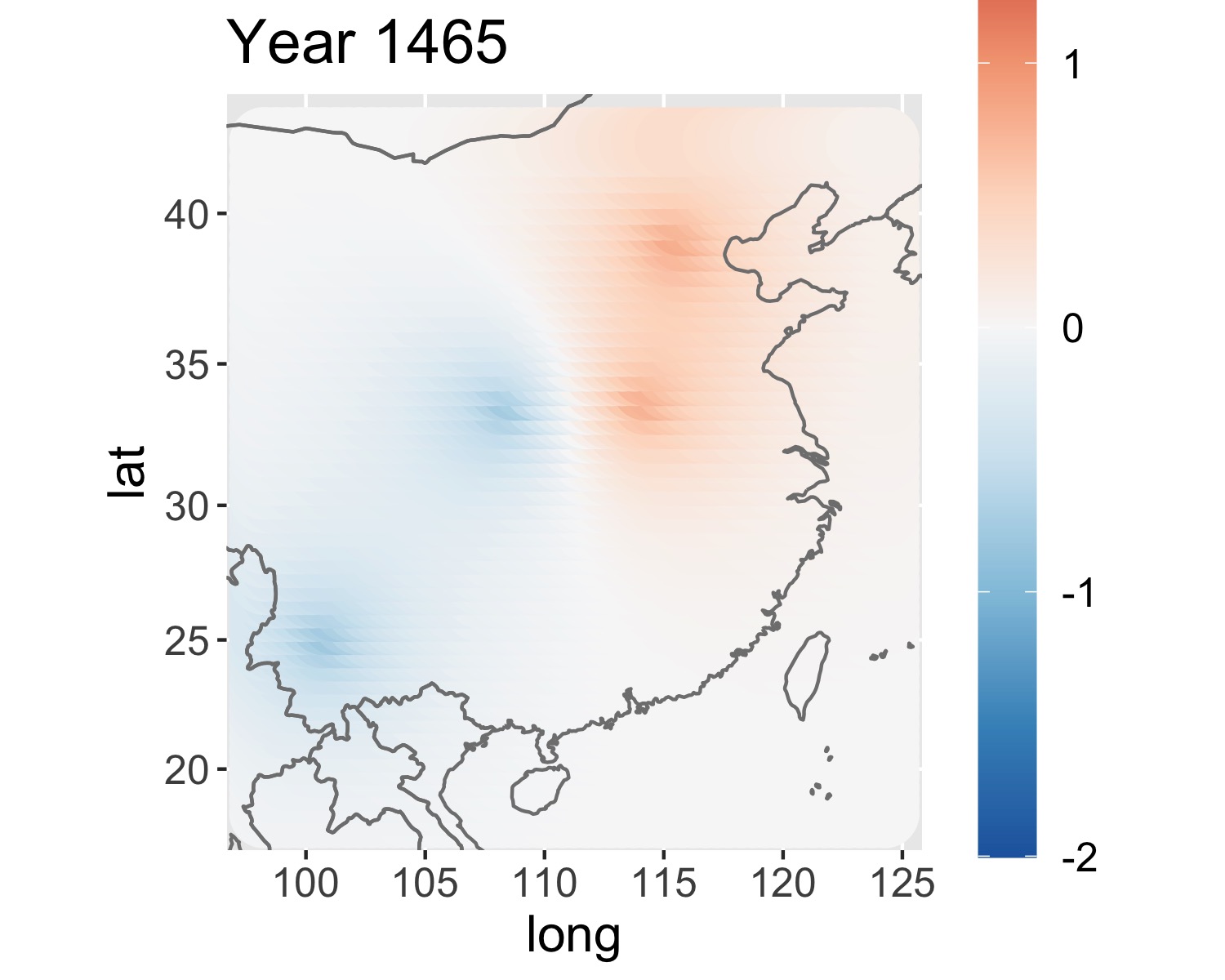}
        \caption{}
    \end{subfigure}
    \caption{(a) The REACHES temperature data for the year 1465; (b) The predicted surface obtained from the best linear predictor based on the data in (a).}
    \label{bakrig1}
\end{figure}

\begin{figure}[tbh]\centering
    \begin{subfigure}{0.49\linewidth}
        \includegraphics[width=\textwidth]{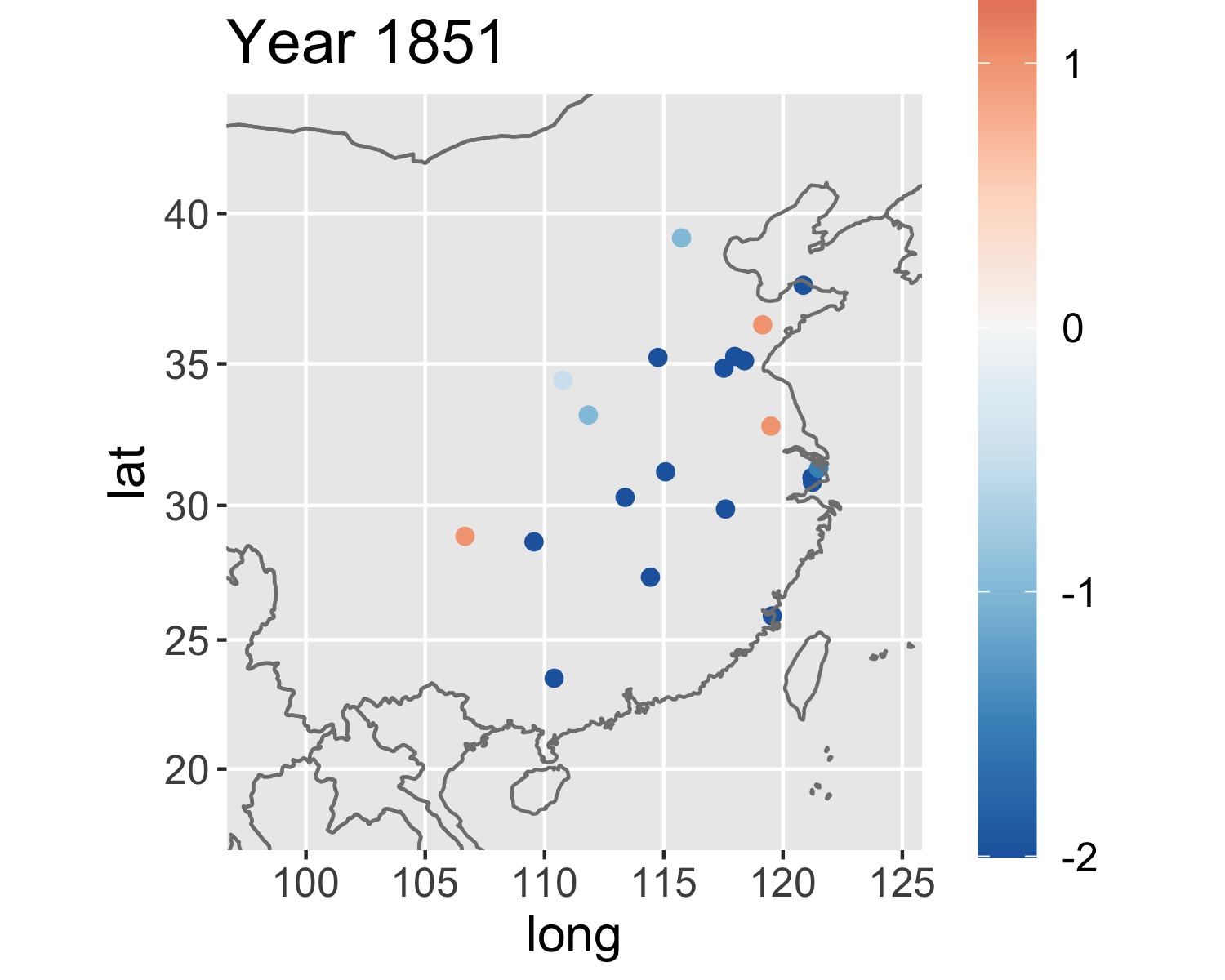}
        \caption{}
    \end{subfigure}
    \hfill
    \begin{subfigure}{0.49\linewidth}
        \includegraphics[width=\textwidth]{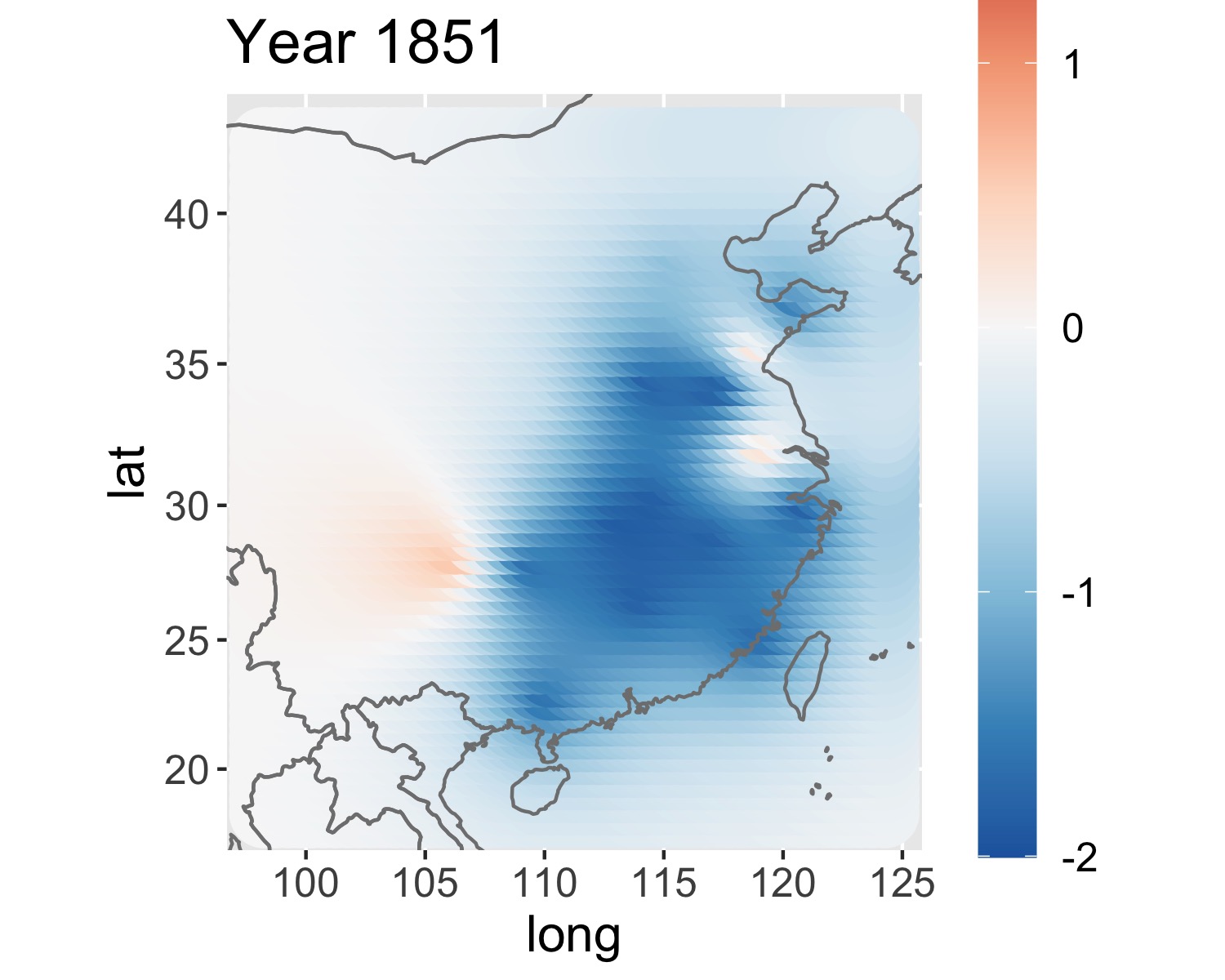}
        \caption{}
    \end{subfigure}
    \caption{(a) The REACHES temperature data for the year 1851; (b) The predicted surface obtained from the best linear predictor based on the data in (a).}
    \label{bakrig}
\end{figure}

\begin{figure}[bt]
    \centering
    \begin{subfigure}{0.48\linewidth}
        \includegraphics[width=\textwidth]{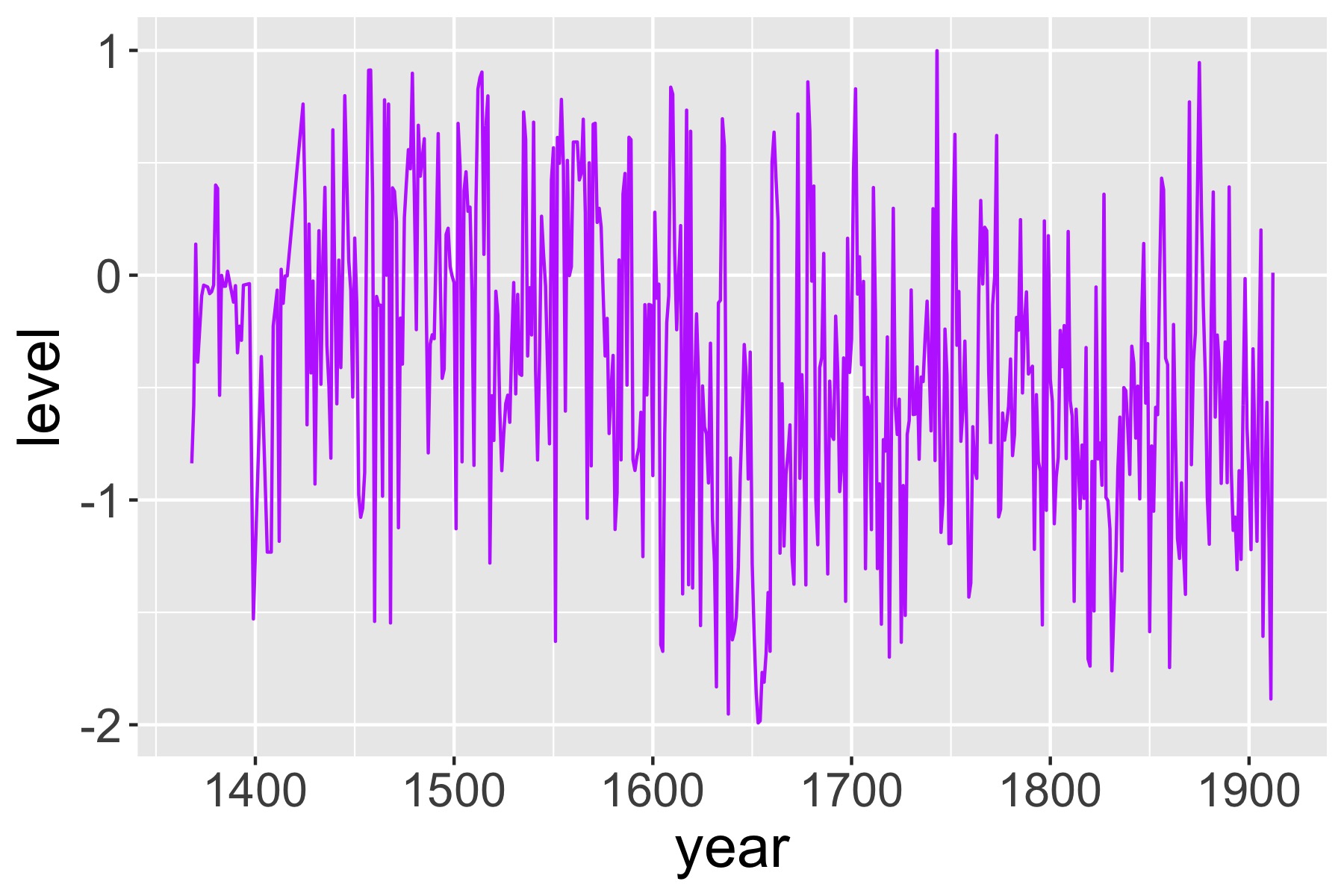}
        \caption{}
    \end{subfigure}
    \hfill
    \begin{subfigure}{0.48\linewidth}
        \includegraphics[width=\textwidth]{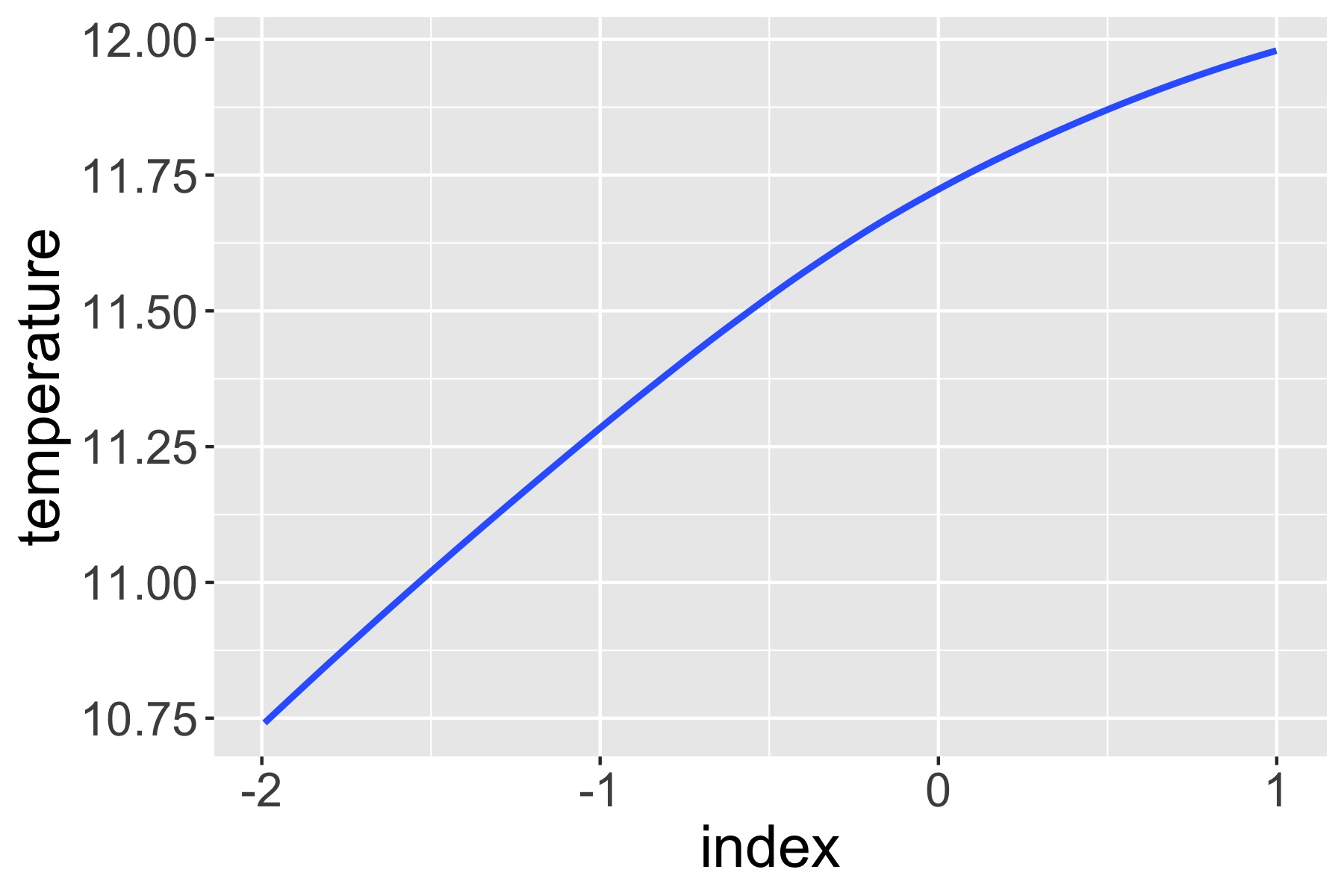}
        \caption{}
    \end{subfigure}
    \caption{(a) Annual temperature time series for Beijing based on kriged REACHES data; (b) Estimated calibration function transforming REACHES temperature indices in Beijing to the Celsius scale.}
    \label{pdftime}
\end{figure}

Figure \ref{Monte carlo}(a)-(c) shows the calibration functions, \(f_1(\cdot)\), \(f_2(\cdot)\), and \(f_3(\cdot)\), respectively. The initial estimates of the parameters \((\alpha, \sigma_Y^2, \sigma_\varepsilon^2)'\) are given by \(\big(\tilde{\alpha}, \tilde{\sigma}_Y^2, \tilde{\sigma}_\varepsilon^2\big)' = (426.7, 0.564, 0.197)\), and the final estimates, after applying bias correction, are \(\big(\hat{\alpha}, \hat{\sigma}_Y^2, \hat{\sigma}_\varepsilon^2\big)' = (299.1, 0.739, 0.146)\). Figures~\ref{bakrig1} and \ref{bakrig} display the original REACHES temperature data alongside the surfaces predicted using the best linear predictor (after substituting the final parameter estimates \(\big(\hat{\alpha}, \hat{\sigma}_Y^2, \hat{\sigma}_\varepsilon^2\big)'\)) for the years 1465 and 1851, respectively. Notice that if we apply ordinary kriging without enforcing a zero mean for $Y(\boldsymbol{s})$, the predicted surface in Figure~\ref{bakrig}(b) would become all negative. Figure~\ref{pdftime}\,(a) presents the annual temperature time series for Beijing, derived from the best linear predictions.

\subsection{Quantile Mapping to Convert Kriged REACHES Data to Celsius Scale}

Although the kriged REACHES surfaces provide historical temperature estimates at any location from 1368 to 1911, these estimates are not expressed on the Celsius scale. To convert the kriged REACHES data into Celsius temperatures, we utilize information from the LME data for the same period by applying quantile mapping.

Let \(\hat{Y}_1, \dots, \hat{Y}_N\) represent the kriged REACHES temperature data at a specific location from 1368 to 1911, which we assume are identically distributed with a continuous cumulative distribution function (CDF) \(F_{\hat{Y}}\). Similarly, let \(x_1, \dots, x_N\) denote the corresponding averaged LME temperature data over the 13 simulation outputs at the same location, assumed to be identically distributed with a continuous CDF \(F_x\). Since \(F_{\hat{Y}}(\hat{Y}_t)\) and \(F_x(x_t)\) are uniformly distributed on \((0,1)\) for \(t = 1, \dots, N\), we have
\begin{equation}
    F_x^{-1}\big( F_{\hat{Y}}( \hat{Y}_t ) \big) \sim F_x, \quad t = 1, \dots, N.
    \label{3.4}
\end{equation}

\noindent This mapping effectively transforms the REACHES data \(\hat{Y}_t\) into the Celsius scale by matching the quantiles of the REACHES data distribution to those of the LME data distribution.

Based on exploratory data analysis, we assume that the REACHES temperature data \(\big\{\hat{Y}_t\big\}\) follow a normal distribution, whereas the LME temperature data \(\{x_t\}\) follow a skew-normal distribution. We apply maximum likelihood (ML) estimation to determine the parameters of these distributions, obtaining the estimated CDFs \(\hat{F}_{\hat{Y}}\) and \(\hat{F}_x\). This leads to the calibration function $g(\cdot)$, which is described as follows:
\begin{equation}
g(y) = \hat{F}_x^{-1}\big( \hat{F}_{\hat{Y}}( y ) \big).
\label{eq:quantile mapping}
\end{equation}

\begin{figure}[tb]\centering
    \begin{subfigure}{0.48\linewidth}
        \includegraphics[width=\textwidth]{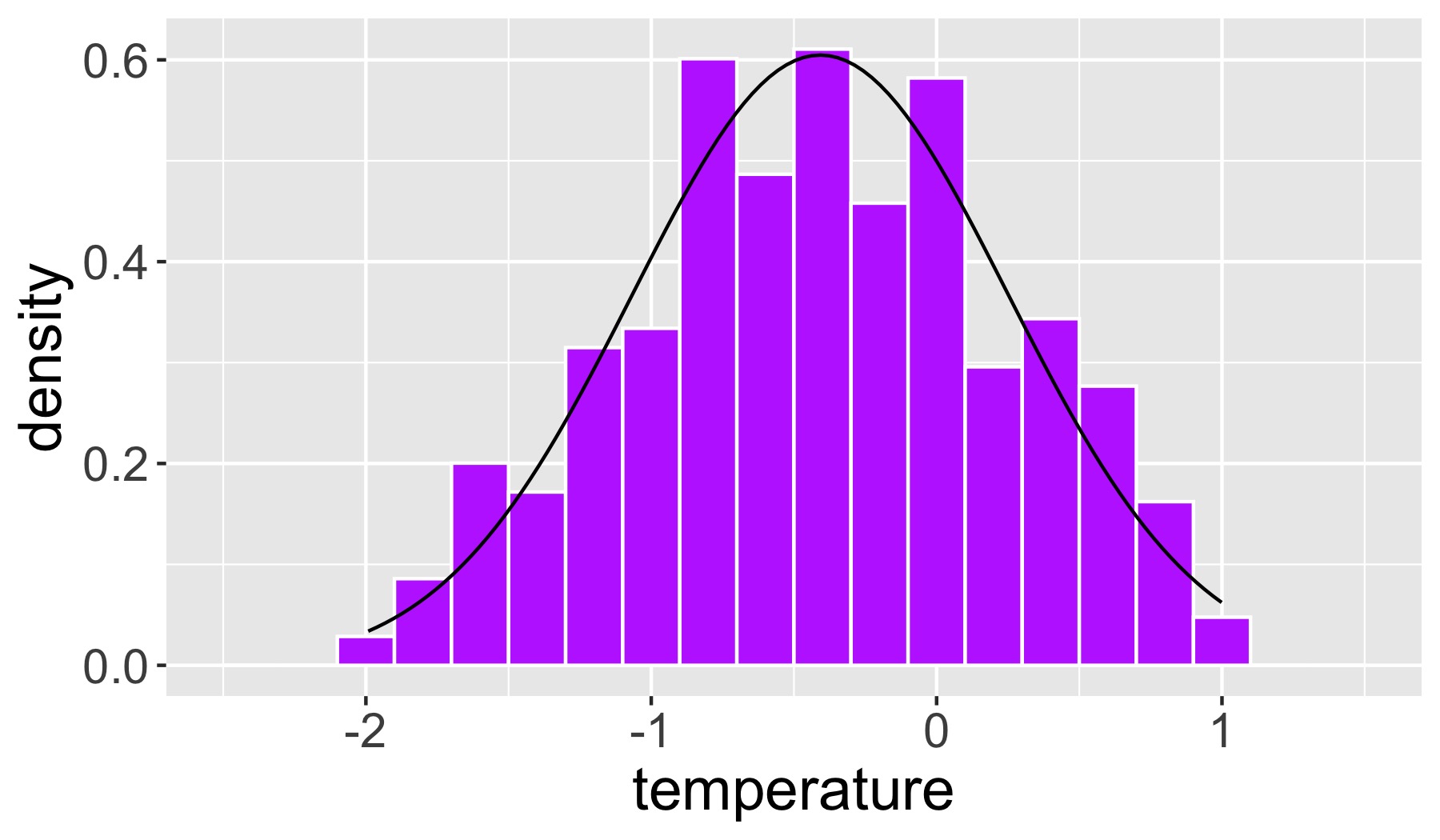}
        \caption{}
    \end{subfigure}
    \hfill
    \begin{subfigure}{0.48\linewidth}
        \includegraphics[width=\textwidth]{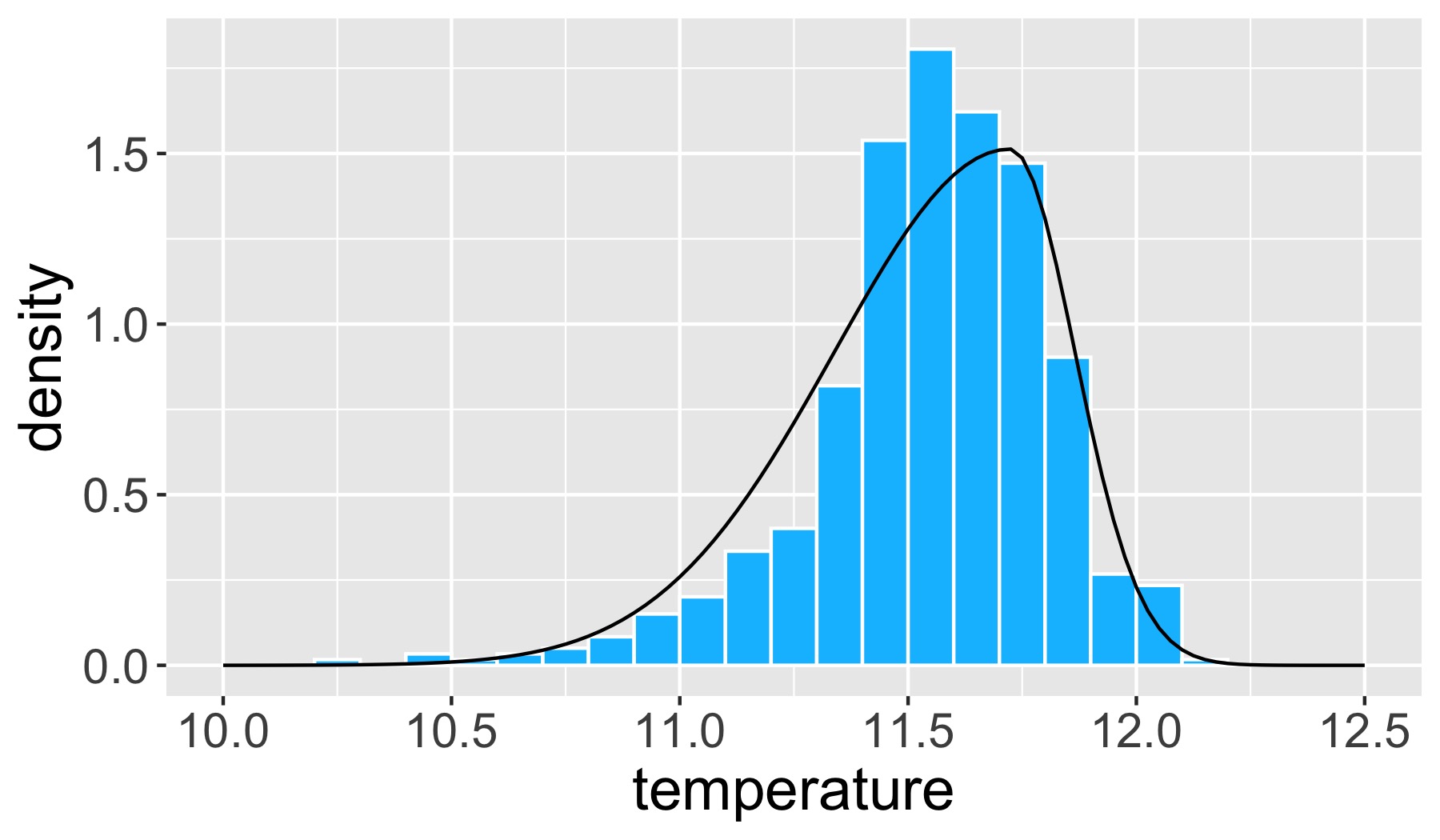}
        \caption{}
    \end{subfigure}
    \caption{Histogram of Beijing temperature data and the corresponding PDF estimate: (a) REACHES data with estimated normal PDF; (b) LME data with estimated skew normal PDF.}
    \label{pdfhist}
\end{figure}

\begin{figure}[tb]\centering
\includegraphics[width=0.6\textwidth]{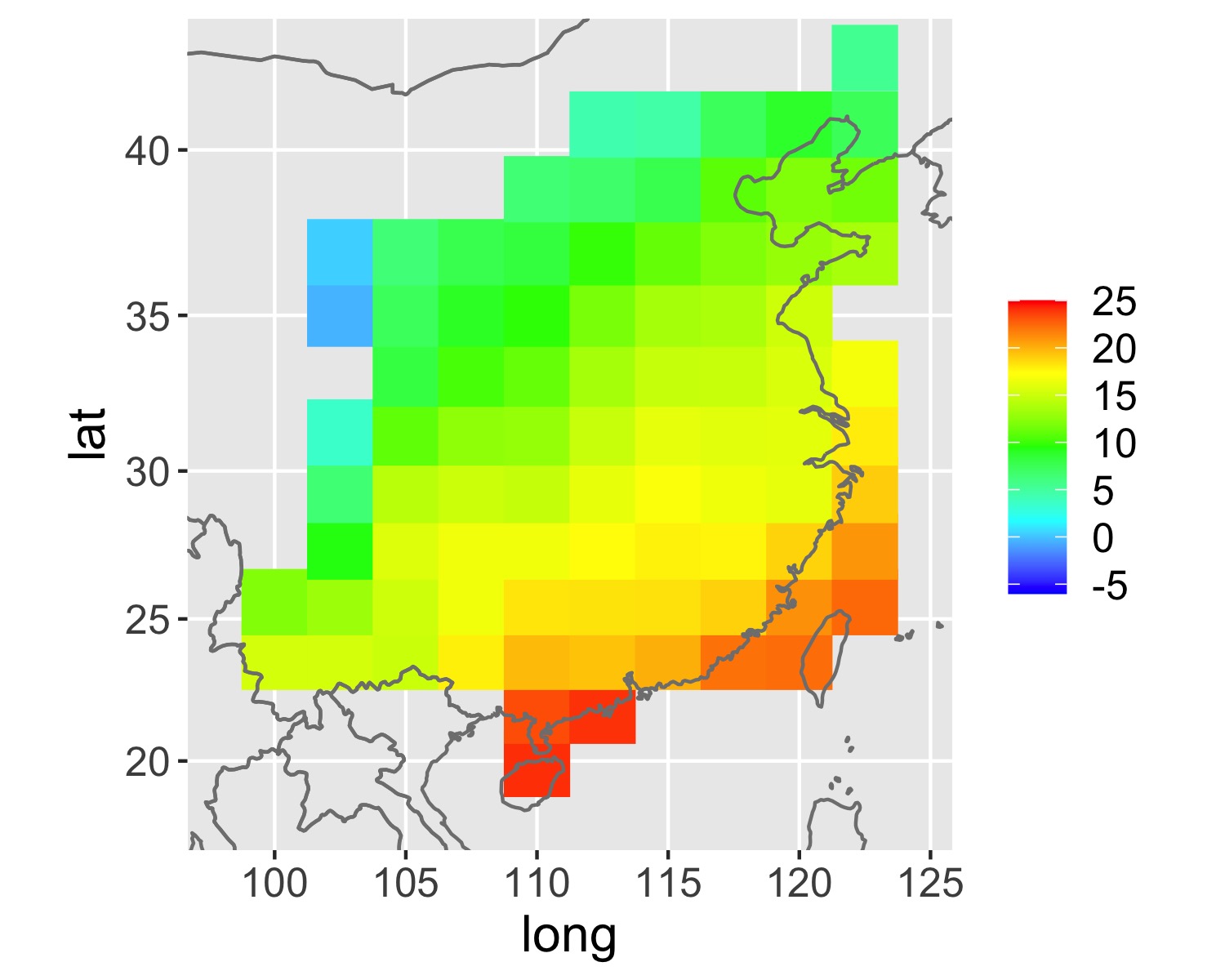}
\caption{A Celsius-scale map corresponding to a REACHES index of $0$.}
\label{value0}
\end{figure}

\noindent The estimated probability density functions (PDFs) are illustrated in Figure~\ref{pdfhist}, and the resulting calibration function that converts REACHES indices to the Celsius scale is shown in Figure~\ref{pdftime}(b). Since REACHES index values are relative measures, they can represent different absolute temperatures at various locations. Figure~\ref{value0} presents a Celsius-scale map corresponding to a REACHES index of \(0\). 

Recall that the MSPE for the kriging predictor \(\hat{Y}_t\) of \(Y_t\) is given by equation~\eqref{eq:MSPE}. Applying a Taylor expansion to \(g(\hat{Y}_t)\) around \(Y_t\), the MSPE of \(g(\hat{Y}_t)\) is approximately:
\begin{equation}
\mathrm{E}\big[ g(\hat{Y}_t) - g(Y_t) \big]^2 \approx \big[ g'(\hat{Y}_t) \big]^2 \, \mathrm{E}\big( \hat{Y}_t - Y_t \big)^2 = \left( \frac{ \hat{f}_{\hat{Y}}( \hat{Y}_t ) }{ \hat{f}_x\big( g( \hat{Y}_t ) \big) } \right)^2 \mathrm{E}\big( \hat{Y}_t - Y_t \big)^2,
\label{eq:MSPE2}
\end{equation}

\noindent where \(\hat{f}_{\hat{Y}}(y) \equiv \hat{F}'_{\hat{Y}}(y)\) and \(\hat{f}_x(y) \equiv \hat{F}'_x(y)\) are the estimated PDFs of \(\hat{Y}_t\) and \(x_t\), respectively. Thus, we obtain a noisy time series, $X^*_t\equiv  g(\hat{Y}_t)$; $t=1,\dots.N$, which can be approximately given by
\begin{equation}
X^*_t=X_t+\delta_t;\quad t=1,\dots,N.
\label{eq:data2}
\end{equation}

\noindent where $X_t\equiv g(Y_t)$ is the true Celsius-scale temperature and $\delta_t\sim N(0,v_t^2)$ is the corresponding error, independent of $X_t$; $t=1,\dots,N$.

\begin{figure}[bpt]
    \centering
    \begin{subfigure}{0.75\linewidth}
        \includegraphics[width=\textwidth]{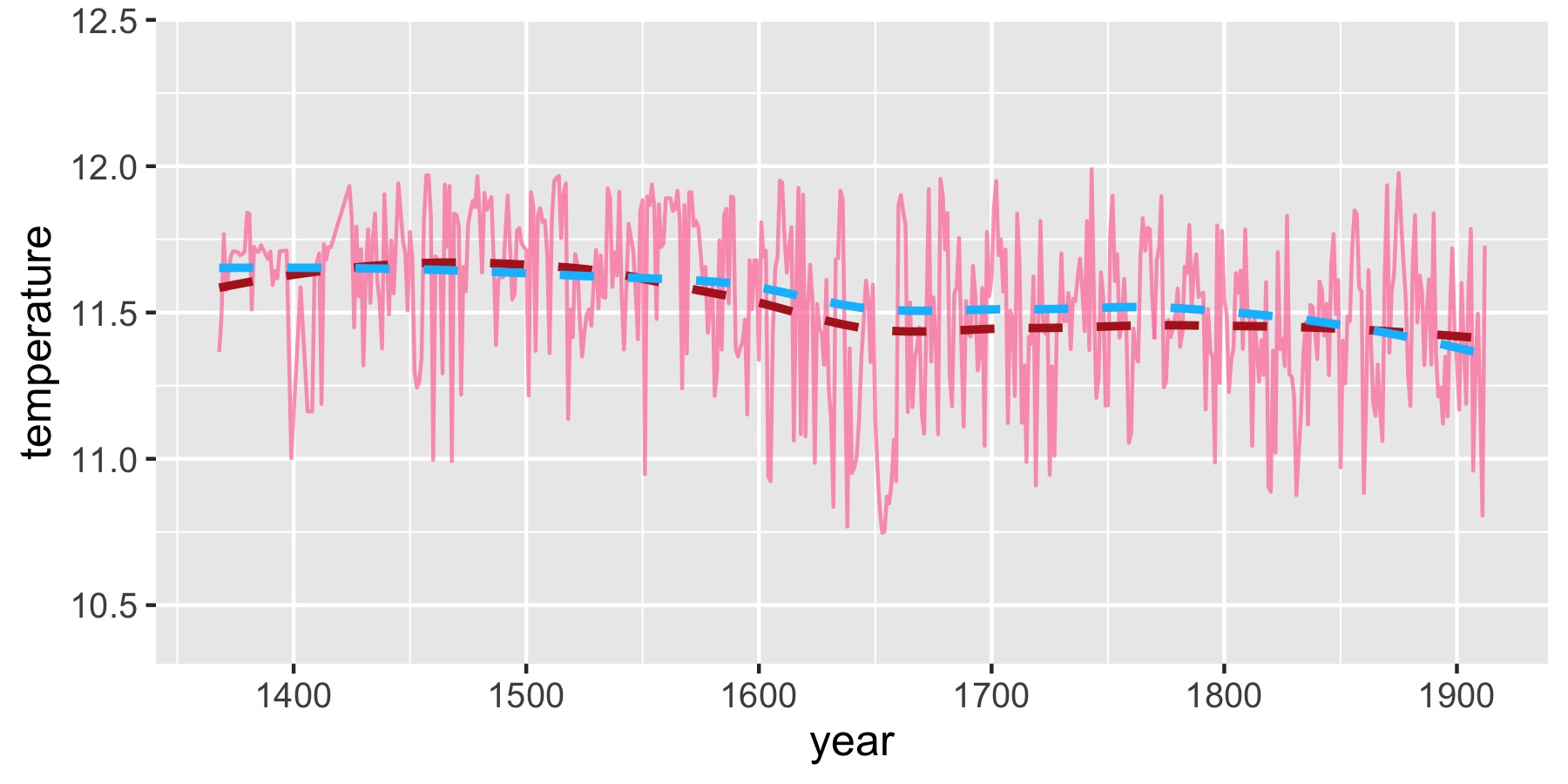}
        \vspace{-0.7cm}
        \caption{}
        \vspace{0.3cm}
    \end{subfigure}
    \vspace{0.3cm}
    \begin{subfigure}{0.75\linewidth}
        \includegraphics[width=\textwidth]{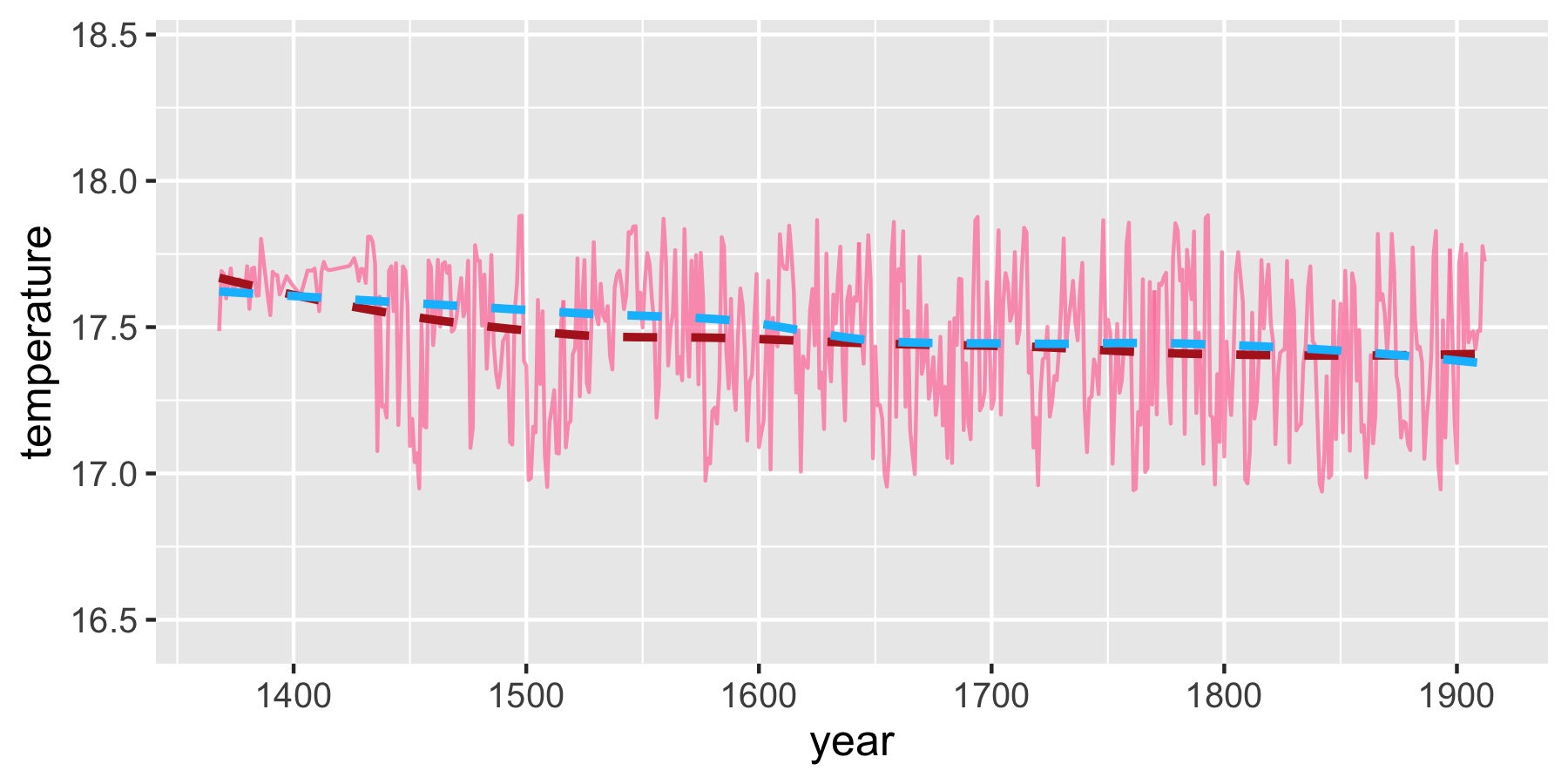}
        \vspace{-0.7cm}
        \caption{}
    \end{subfigure}
    \vspace{0.3cm}
    \begin{subfigure}{0.75\linewidth}
        \includegraphics[width=\textwidth]{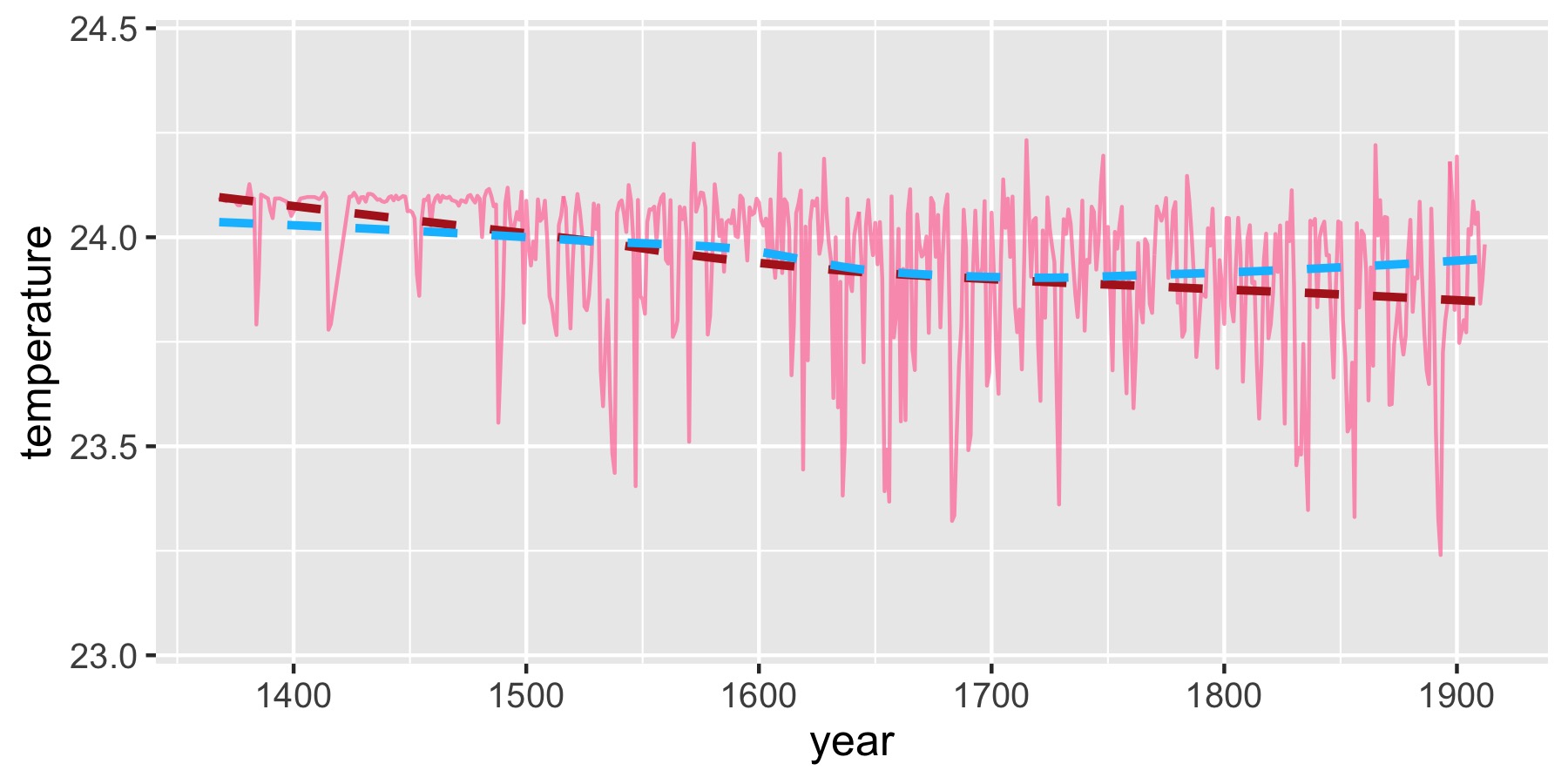}
        \vspace{-0.7cm}
        \caption{}
    \end{subfigure}
    \caption{Time series plots of Celsius-scaled REACHES data from 1368 to 1911. The red and blue dashed curves represent the LOESS fits of the REACHES data and the LME data, respectively: (a) Beijing, (b) Shanghai, and (c) Hong Kong.}
    \label{tme}
\end{figure}

Figure~\ref{tme} presents the Celsius-scale REACHES data of \eqref{eq:data2} for Beijing, Shanghai, and Hong Kong, transformed via the function $g(\cdot)$ as defined in \eqref{eq:quantile mapping}. The LOESS (Locally Estimated Scatterplot Smoothing) curves \citep{Cleveland1979}, shown as blue and red dashed lines for the LME and Celsius-scale REACHES data, respectively, reveal consistent long-term trends between these two fundamentally different sources. This agreement underscores the reliability of the REACHES data in reconstructing historical temperatures.

Moreover, both the REACHES and LME datasets consistently capture significant climatic events, such as the Little Ice Age, which spans from the 14th to the 19th century. This period, noted for global cooling documented in multiple studies \citep{Jones2004, Zheng2014, Fan2023}, is reflected in the cooling trends observed in both the REACHES and LME data. This congruence highlights the value of the REACHES data as an effective climate proxy, corroborating its credibility in climate reconstruction studies.

\section{Bayesian Data Assimilation}\label{sec4}

Since the REACHES dataset is constructed from historical documents, its accuracy is highly uncertain. To enhance the precision of our climate reconstruction, we employ a Bayesian data assimilation framework that integrates both the REACHES data and the LME temperature simulations. This framework has been widely applied across various scientific disciplines to combine observational data with model-based predictions, effectively accounting for uncertainties in both sources \citep{WIKLE2007}. In this study, we focus on Beijing, Shanghai, and Hong Kong to illustrate our analysis, as these cities have the largest number of temperature records prior to 1912.

We propose a nonstationary first-order autoregressive model (see, e.g., \citealt{tsay2005}) to form the prior temperature. We use the LME data to estimate the model parameters and treat the resulting distribution as the prior. We then integrate the Celsius-scale REACHES data to estimate the posterior distribution using the Kalman filter and smoother algorithms.

\subsection{Prior Derivation}

We assume that the state process follows a known transition model \( p\big(X^{(j)}_{t} | X^{(j)}_{t-1}\big) \), which can be expressed as:
\begin{align}
    X^{(j)}_{t} - \mu_t
=&~ M_{t-1} \big(X^{(j)}_{t-1} - \mu_{t-1} \big) + \eta_t, \quad \eta^{(j)}_t \sim N(0, r_t^2);\quad t=2,\dots,N,
\label{eq:prior1}\\
    X^{(j)}_1
\sim&~ N(\mu_1, r_1^2),
\label{eq:prior2}
\end{align}

\noindent for $j=1,\dots,13$ corresponding to 13 LME simulation outputs, where \( X^{(j)}_t \) denotes the temperature at time \( t \) in the \( j \)-th simulation, \( \mu_t = \mathrm{E}\big(X^{(j)}_t\big) \) is the expected value of \( X^{(j)}_t \), \( M_{t-1} \) is an AR parameter describing the temporal evolution of the process, and \( \eta^{(j)}_t \) represents the process noise, which is assumed to be independent of \( X^{(j)}_{t-1} \), for $t=1,\dots,N$ and $j=1,\dots,13$.

The model given by \eqref{eq:prior1} and \eqref{eq:prior2} describes how the temperature state evolves over time based on the LME simulations. By estimating the parameters from the LME data, we create a prior distribution that reflects both the physical dynamics illustrated by the climate model and the variability in the simulations.

Let \( \boldsymbol{\theta} = (\mu_1, \dots, \mu_N, r_1^2, \dots, r_N^2, M_1, \dots, M_{N-1})' \) denote the collection of parameters in the model given by \eqref{eq:prior1} and \eqref{eq:prior2}. To estimate \( \boldsymbol{\theta} \), we employ the LME data using penalized ML with a fused lasso penalty, encouraging both sparsity and smoothness in the parameter estimates.

We define \( \boldsymbol{X}^{(j)} = \big(X_1^{(j)}, \dots, X_N^{(j)}\big)' \) as the \( j \)-th time series and organize all LME simulations into a matrix \( \boldsymbol{X} = (\boldsymbol{X}^{(1)}, \dots, \boldsymbol{X}^{(13)}) \). Under \eqref{eq:prior1} and \eqref{eq:prior2}, the negative log-likelihood function for the \( j \)-th time series is given by:
\begin{equation*}
  h(\boldsymbol{X}^{(j)};\boldsymbol{\theta})
= \frac{1}{2}\sum_{t=1}^N \log(r_t^2) + \frac{\big({X}_{1}^{(j)}-\mu_1\big)^2}{2 {r}_1^2} + \sum_{t=2}^N\frac{\big(X_t^{(j)} - {\mu}_t- {M}_{t-1} (X_{t-1}^{(j)} - {\mu}_{t-1})\big)^2}{2 r_t^2}+\mbox{constant}.
\end{equation*}

\noindent Aggregating over all simulations while promoting smoothness among neighboring parameters, we consider the negative log-likelihood with a fused lasso penalty:
\begin{equation}
g(\boldsymbol{X}; \boldsymbol{\theta}) = \sum_{j=1}^{13}h(\boldsymbol{X}; \boldsymbol{\theta}) + \lambda_1 \sum_{t=1}^{N-1} |M_t| + \lambda_2 \sum_{t=1}^{N-2} |M_{t+1} - M_t| + \lambda_3 \sum_{t=1}^{N-1} |\mu_{t+1} - \mu_t|,
\label{eq:PML}
\end{equation}

\noindent where \( \lambda_1\geq 0 \), \( \lambda_2\geq 0 \), and \( \lambda_3\geq 0 \) are tuning parameters controlling the degree of regularization. The first penalty term encourages sparsity in the $\{M_t\}$ coefficients, the second promotes smoothness across time for the $\{M_t\}$, and the third enforces smoothness in the mean parameters $\{\mu_t\}$. Our goal is to estimate the parameter vector \( \boldsymbol{\theta} \) by
\[
\boldsymbol{\hat{\theta}}_{\lambda_1, \lambda_2, \lambda_3} = \arg\min_{\boldsymbol{\theta}} \, g(\boldsymbol{X}; \boldsymbol{\theta}).
\]

Including the fused lasso penalty serves to regularize the parameter estimates, mitigating overfitting and ensuring that the estimated parameters exhibit temporal smoothness, a reasonable assumption for climate variables that change gradually over time. However, due to the absence of closed-form solutions for \( \boldsymbol{\hat{\theta}}_{\lambda_1, \lambda_2, \lambda_3} \), we employ an iterative algorithm to estimate the parameters. Specifically, we use the R package \texttt{penalized} \citep{Goeman2022}, which implements efficient coordinate descent methods suitable for high-dimensional optimization problems with lasso-type penalties. The detailed steps are given below:
\begin{enumerate}
    \item Initialization: $\mu_t = \displaystyle\frac{1}{13} \sum_{j=1}^{13} X_t^{(j)}$ and $r_t^2 = \displaystyle\frac{1}{13} \sum_{j=1}^{13} \big( X_t^{(j)} - \mu_t \big)^2$, for $t = 1, \dots, N$.
    \item Estimate $M_t$: With \( \mu_t \) and \( r_t^2 \) fixed, estimate \( M_t \) by minimizing \( g(\boldsymbol{X}; \boldsymbol{\theta}) \) with respect to \( M_t \), using the \texttt{penalized} package to handle a fused lasso penalty.
    \item Update $\mu_t$: With \( r_t^2 \) and \( M_t \) fixed, estimate \( \mu_t \) by minimizing \( g(\boldsymbol{X}; \boldsymbol{\theta}) \) with respect to \( \mu_t \), using the \texttt{penalized} package.
    \item Update $r_t^2$: With \( \mu_t \) and \( M_t \) fixed, update \( r_t^2 \) using the closed-form solution:
            \begin{align*}
            r_1^2 &= \frac{1}{13} \sum_{j=1}^{13} \big( X_1^{(j)} - \mu_1 \big)^2, \\
            r_t^2 &= \frac{1}{13} \sum_{j=1}^{13} \big\{ X_t^{(j)} - \mu_t - M_{t-1} \big( X_{t-1}^{(j)} - \mu_{t-1} \big) \big\}^2, \quad \text{for } t = 2, \dots, N.
            \end{align*}
       \item Repeat steps 2-4 until convergence.
\end{enumerate}

The choice of tuning parameters \(\lambda_1, \lambda_2,\) and \(\lambda_3\) is crucial in the estimation process. When \(\lambda_1 = \lambda_2 = \lambda_3 = 0\), the method reduces to the standard ML estimation. As \(\lambda_1 \rightarrow \infty\), the estimates of the coefficients \(M_t\) shrink toward zero, promoting sparsity in the model. Conversely, as \(\lambda_2 \rightarrow \infty\) and \(\lambda_3 \rightarrow \infty\), the estimates of \(M_t\) and \(\mu_t\) converge to constant values corresponding to a stationary time series, enhancing smoothness over time by reducing variability among adjacent parameters.

To determine the optimal tuning parameters, we perform 13-fold cross-validation, where each fold contains a single LME time series. We treat the \(l\)-th time series as the testing data in each iteration and use the remaining 12 as the training data. Let \(\boldsymbol{\hat{\theta}}_{\lambda_1, \lambda_2, \lambda_3}^{(-l)}\) denote the estimated parameters based on the training data obtained by solving:
\[
\boldsymbol{\hat{\theta}}_{\lambda_1, \lambda_2, \lambda_3}^{(-l)} = \arg\min_{\boldsymbol{\theta}} \left\{ \sum_{j \ne l} h(\boldsymbol{X}^{(j)};\boldsymbol{\theta}) + \lambda_1 \sum_{t=1}^{N-1} |M_t| + \lambda_2 \sum_{t=1}^{N-2} |M_{t+1} - M_t| + \lambda_3 \sum_{t=1}^{N-1} |\mu_{t+1}-\mu_t| \right\}.
\]
We then compute the cross-validation score:
\[
CV_{\lambda_1, \lambda_2, \lambda_3} = \frac{1}{13} \sum_{l=1}^{13} h\big(\boldsymbol{X}^{(l)};\boldsymbol{\hat{\theta}}_{\lambda_1, \lambda_2, \lambda_3}^{(-l)}\big).
\]
We select the tuning parameters \(\lambda_i > 0\), for \(i = 1, 2, 3\), that minimize \(CV_{\lambda_1, \lambda_2, \lambda_3}\).

\begin{figure}[pbt]
    \centering
    \includegraphics[width=0.85\textwidth]{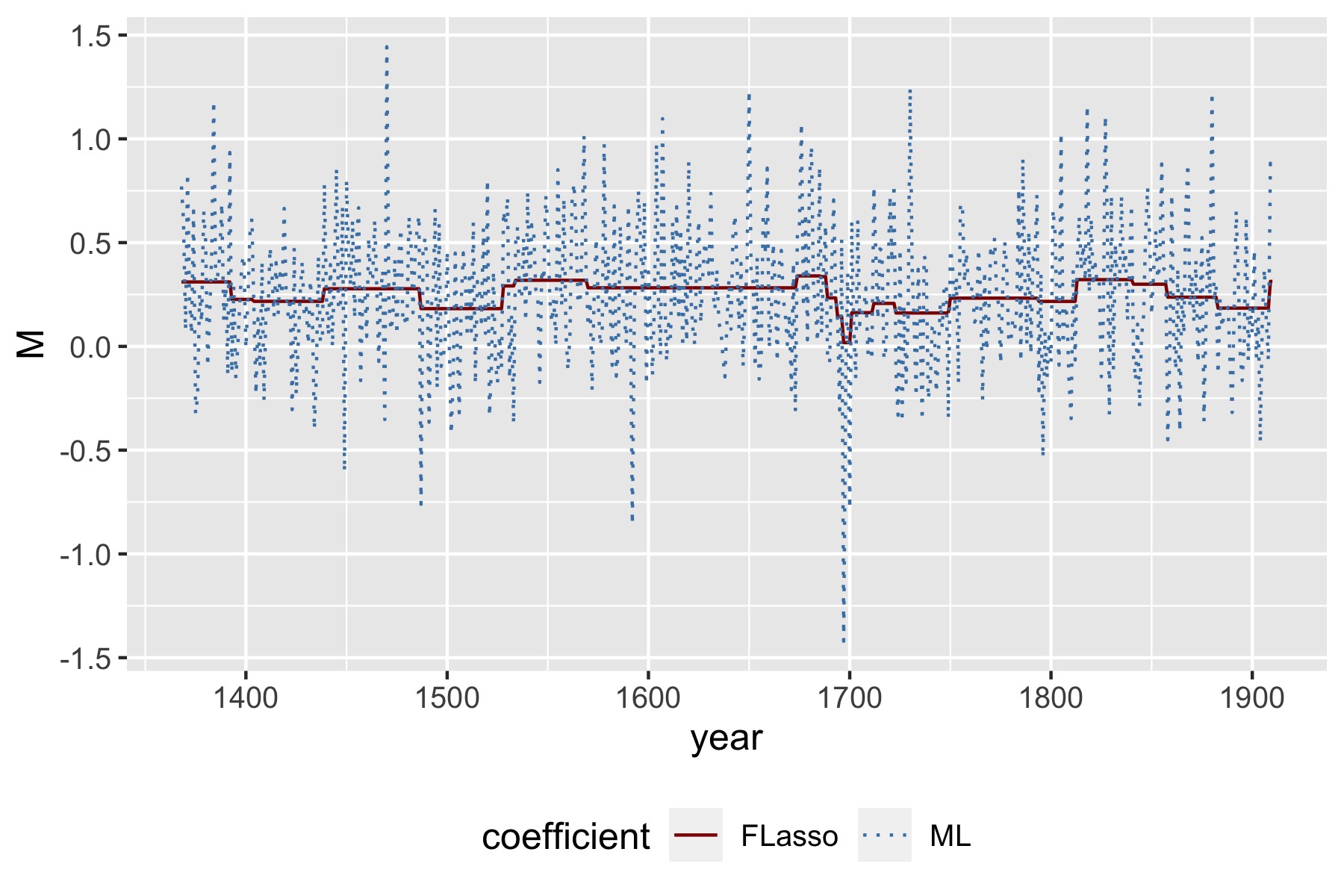}
    \caption{Time series plot of the estimated parameter \(M_t\) from years 1368 to 1911. The fused lasso estimates (FLasso) are in red, and the ML estimates are in blue.}
    \label{mt}
\end{figure}

\begin{figure}[pbt]
    \centering
    \includegraphics[width=0.85\textwidth]{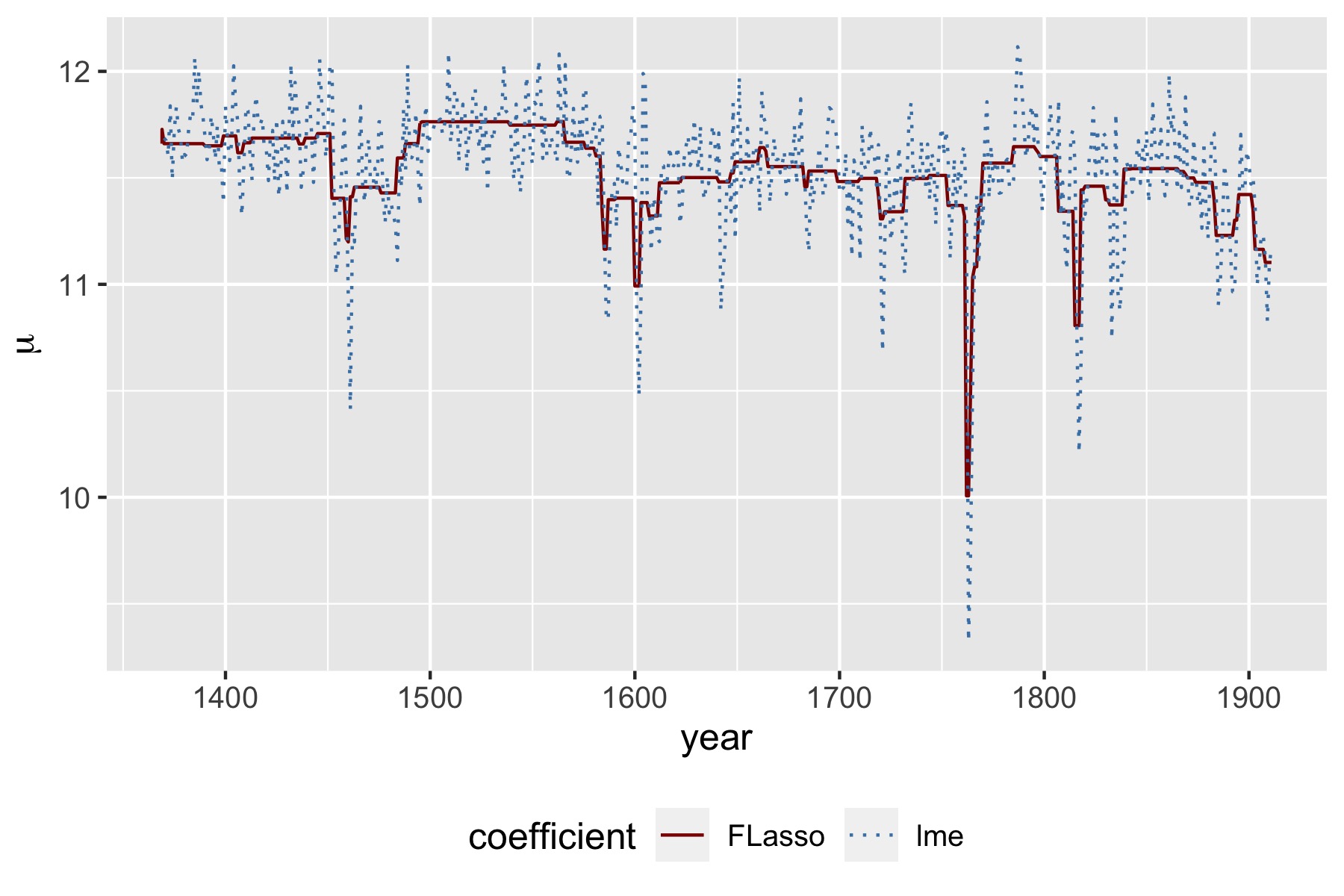}
    \caption{Time series plot of the estimated parameter \(\mu_t\) from years 1368 to 1911. The fused lasso estimates (FLasso) are in red, and the ML estimates are in blue.}
    \label{mu1}
\end{figure}

Figure~\ref{mt} illustrates the estimated \(\{M_t\}\) obtained using the fused lasso penalty, alongside the estimates derived from the ML solution (\(\lambda_1 = \lambda_2 = \lambda_3 = 0\)). Notably, the fused lasso estimates exhibit pronounced smoothness over time, as adjacent \(M_t\) values are more similar, and some coefficients are shrunk toward constants. Figure~\ref{mu1} shows the estimated \(\{\mu_t\}\) employing the fused lasso penalty, along with the ML solution (the yearly average of the 13 LME time series). The plot shows smooth transitions observed among adjacent \(\mu_t\) values when employing the fused lasso penalty, indicating that the estimated mean temperatures vary gradually over time.

\subsection{Posterior Derivation}

Our objective is to estimate the posterior distribution of the true temperature state \( X_t \) given all observations up to time \( N \), that is, \( p(X_t | X_1^*, \dots, X_N^*) \) for \( t = 1, \dots, N \), where \( X_t^* \) denotes the Celsius-scale REACHES observations as defined in \eqref{eq:data2}. We aim to compute the posterior mean \( X_{t|N} \equiv \mathrm{E}(X_t | X_1^*, \dots, X_N^*) \) and the posterior variance \( P_{t|N} \equiv \mathrm{var}(X_t | X_1^*, \dots, X_N^*) \), for $t=1,\dots,N$, based on the nonstationary time series model described in \eqref{eq:prior1} amd \eqref{eq:prior2}.

To achieve this, we use the Kalman filter and Kalman smoother algorithms. Our approach falls under offline (or batch) data assimilation techniques, as we process the entire dataset collectively rather than assimilating data sequentially in real time.

The Kalman filter operates in two steps: the prediction step and the update step. In the prediction step, we use the prior model to project the state estimate forward in time, calculating the prior estimates \( X_{t|t-1} \) and their uncertainties \( P_{t|t-1} \):
\begin{align}
X_{t|t-1} =&~ \mu_t + M_{t-1} ( X_{t-1|t-1} - \mu_{t-1} ), \label{prediction_mean} \\
P_{t|t-1} =&~ M_{t-1}^2 P_{t-1|t-1} + r_t^2, \label{prediction_variance}
\end{align}

\noindent where \( X_{t|t-1} \) is the estimate of \( X_t \) given observations up to time \( t-1 \), and \( P_{t|t-1} \) is the associated variance, for $t=2,\dots,N$.

In the update step, we incorporate  \( X_t^* \) to obtain the posterior estimates \( X_{t|t} \) and \( P_{t|t} \) based on the data up to time $t$:
\begin{align}
K_t &= \frac{P_{t|t-1}}{P_{t|t-1} + v_t^2}, \label{kalman_gain} \\
X_{t|t} &= X_{t|t-1} + K_t ( X_t^* - X_{t|t-1} ), \label{update_mean} \\
P_{t|t} &= (1 - K_t) P_{t|t-1}, \label{update_variance}
\end{align}

\noindent where \( K_t \) is the Kalman gain, determining the weighting between $X_{t|t-1}$ and the new observation $X^*_t$, for $t=2,\dots,N$.

After processing all observations using the Kalman filter, we apply the Kalman smoother to refine the state estimates by incorporating information from all observations. The Kalman smoother provides the posterior estimates \( X_{t|N} \) and \( P_{t|N} \) for all \( t = 1, \dots, N \), using both past and future data:
\begin{align}
J_t &= P_{t|t} M_t P_{t+1|t}^{-1}, \label{smoother_gain} \\
X_{t|N} &= X_{t|t} + J_t ( X_{t+1|N} - X_{t+1|t} ), \label{smoother_mean} \\
P_{t|N} &= P_{t|t} + J_t ( P_{t+1|N} - P_{t+1|t} ) J_t', \label{smoother_variance}
\end{align}

\noindent where \( J_t \) is the smoother gain.

\begin{figure}[bt]
    \centering
    \includegraphics[width=0.95\textwidth]{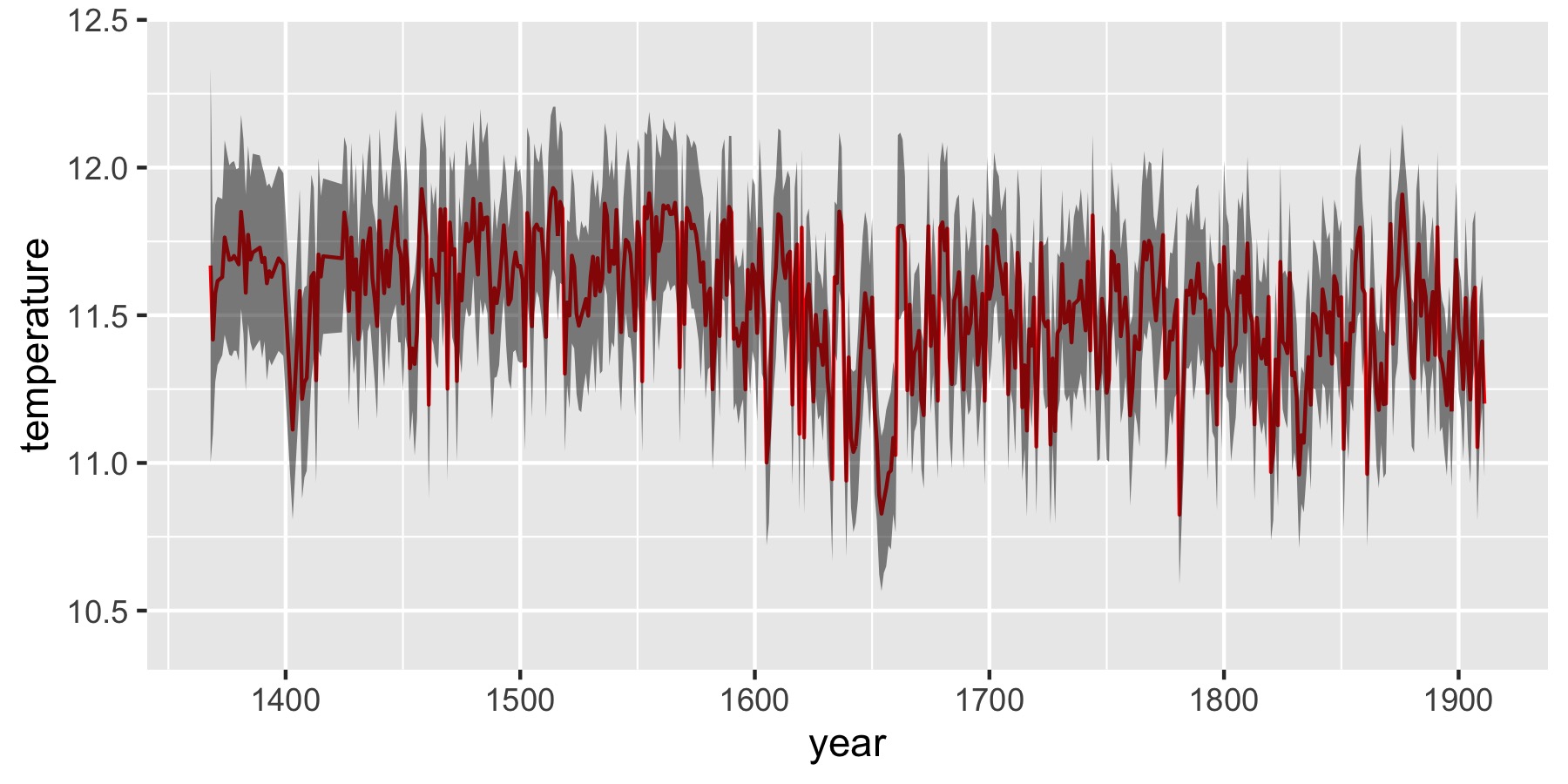}
    \caption{Predicted temperatures in Beijing from 1368 to 1911. The solid line represents the predicted values, and the grey-shaded region indicates the $\pm 1$ standard deviation range.}
    \label{xtt3}
\end{figure}

By integrating the Kalman filter and smoother, we obtain posterior estimates of $\{X_t\}$, leveraging all available observations. Figure~\ref{xtt3} displays the smoothed state estimates \(X_{t|N}\) along with their estimated standard deviations \( P_{t|N}^{1/2} \) for Beijing from 1368 to 1911. Notably, the uncertainty in the estimates has decreased slightly in more recent years due to increased data availability.

This data assimilation method effectively combines the strengths of the LME simulations and the historical observations from REACHES. By systematically updating our state estimates using both prior information and observational data, we achieve a more accurate and reliable reconstruction of historical temperatures.

\section{Assessment of Predicted Temperature Reliability}
\label{sec5}

Assessing the reliability of our predicted temperature data is crucial, especially in the historical context of the Ming and Qing dynasties, where precise instrumental records were largely unavailable prior to the mid-19th century. However, the Global Historical Climatology Network (GHCN) offers temperature records for China starting from the mid-19th century, with observations beginning in Beijing in 1861, Shanghai in 1857, and Hong Kong in 1853. Although there may be inaccuracies associated with early mercury thermometers, these records serve as invaluable benchmarks for validating our predicted temperatures \citep{Bronnimann2019}. Monthly GHCN data can be accessed at \url{https://www.ncei.noaa.gov/products/land-based-station/global-historical-climatology-network-monthly}.

\begin{figure}[p]
    \centering
    \begin{subfigure}{0.32\linewidth}
        \includegraphics[width=\textwidth]{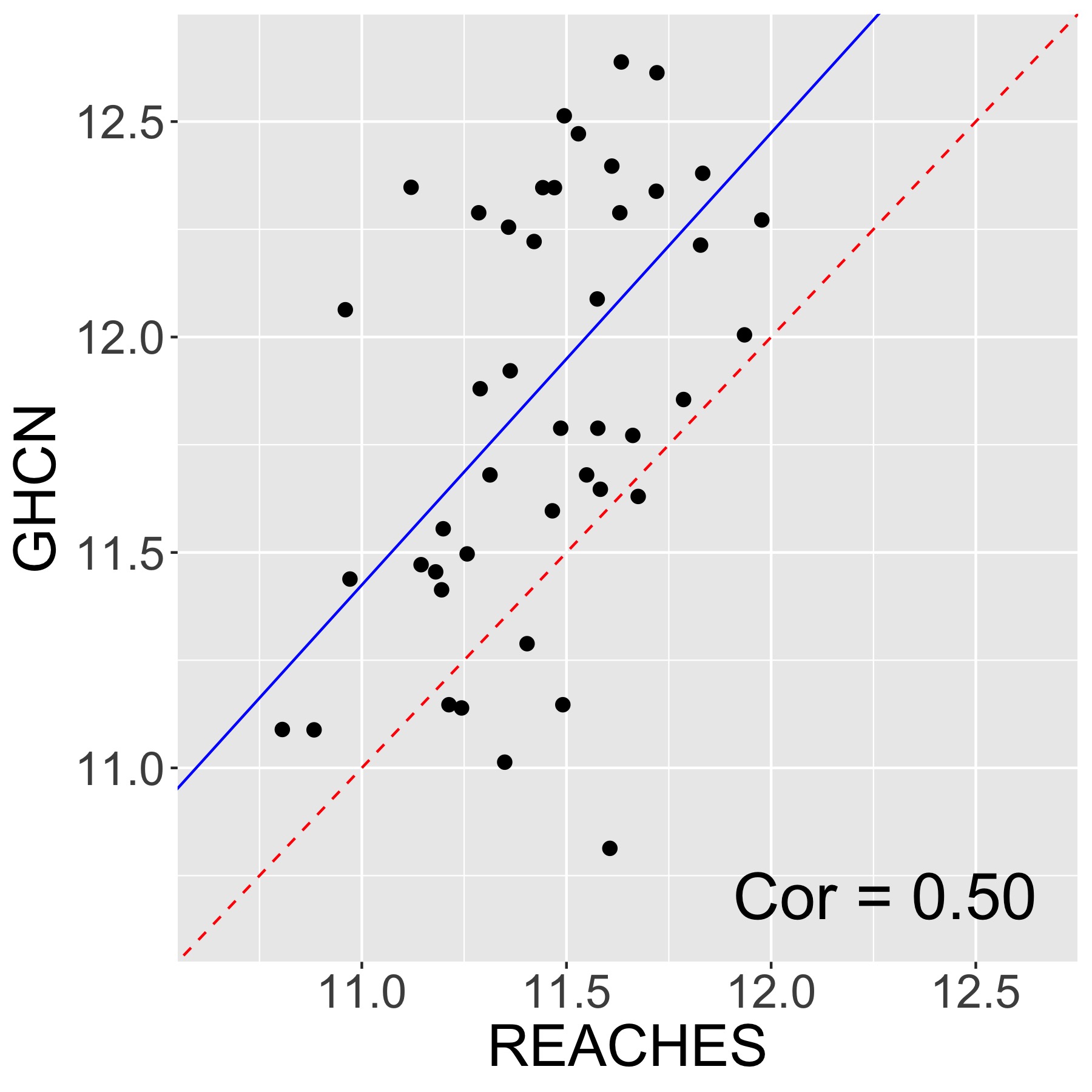}
        \caption{}
    \end{subfigure}
    \hfill
    \begin{subfigure}{0.32\linewidth}
        \includegraphics[width=\textwidth]{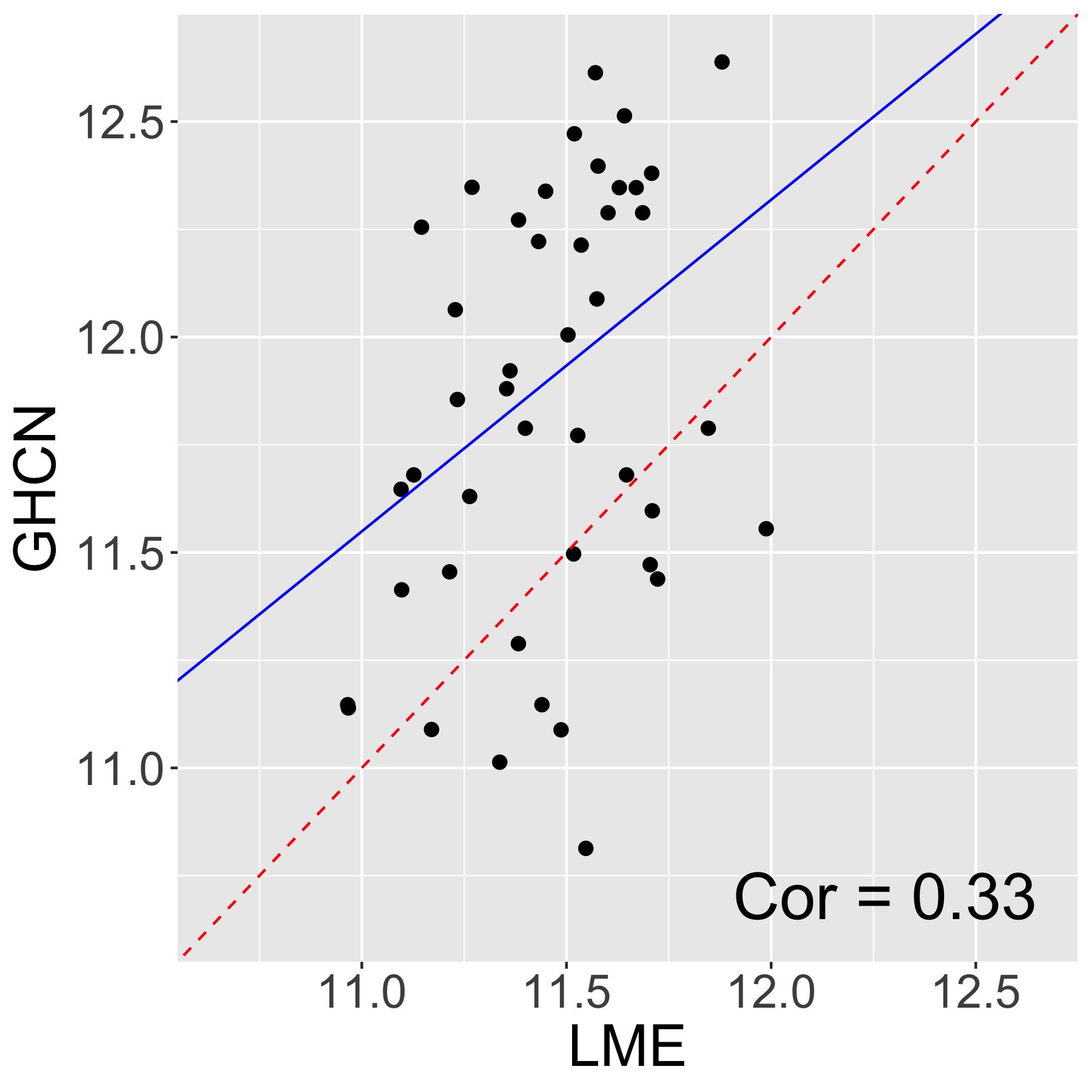}
        \caption{}
    \end{subfigure}
    \hfill
    \begin{subfigure}{0.32\linewidth}
        \includegraphics[width=\textwidth]{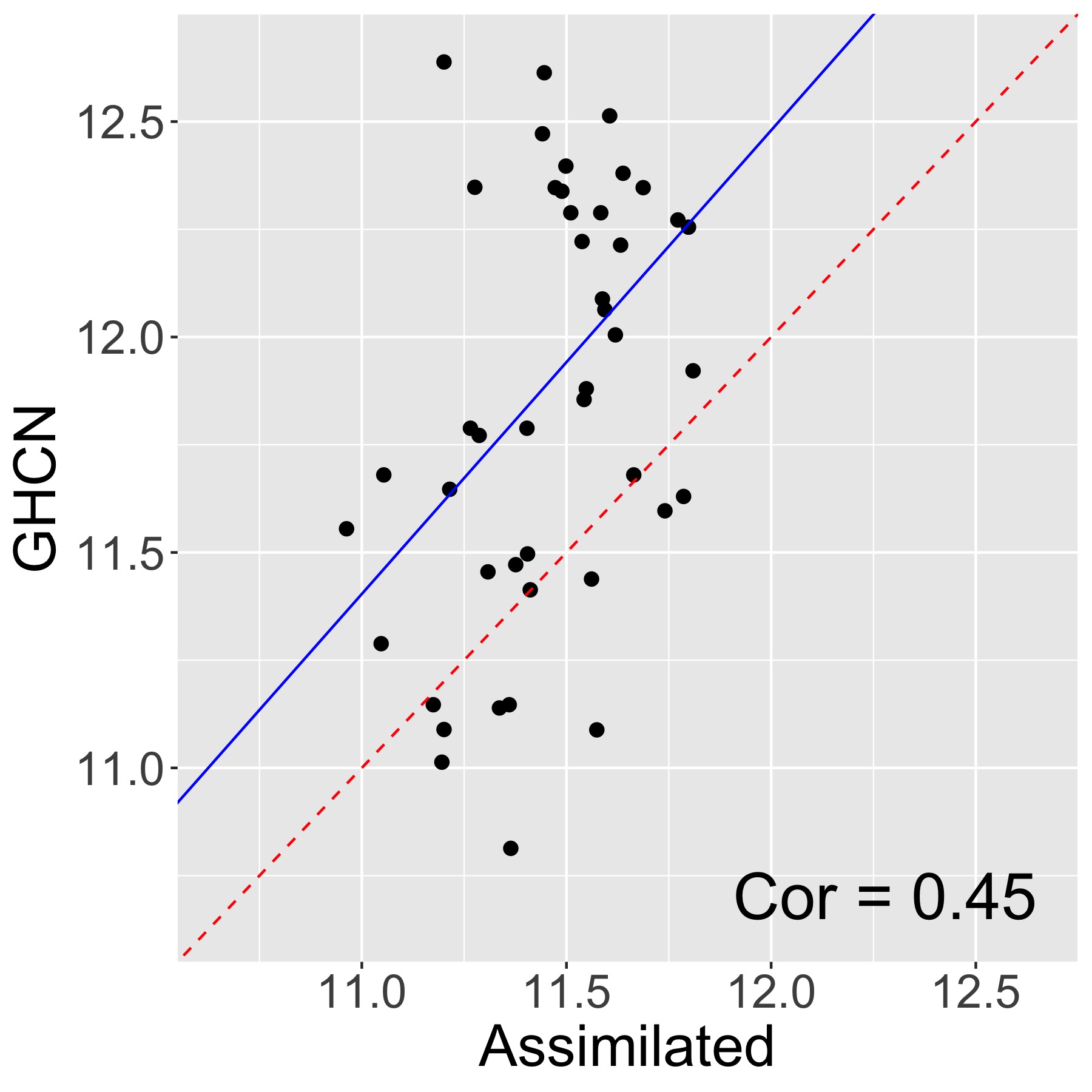}
        \caption{}
    \end{subfigure}
 \caption{Scatter plots between GHCN temperatures and estimates in Beijing from 1861 to 1911: (a) GHCN vs.\ Celsius-scale REACHES; (b) GHCN vs.\ LME; (c) GHCN vs.\ Assimilated.} 
     \label{ghcnn}
\end{figure}

\begin{figure}[p]
    \centering
    \begin{subfigure}{0.32\linewidth}
        \includegraphics[width=\textwidth]{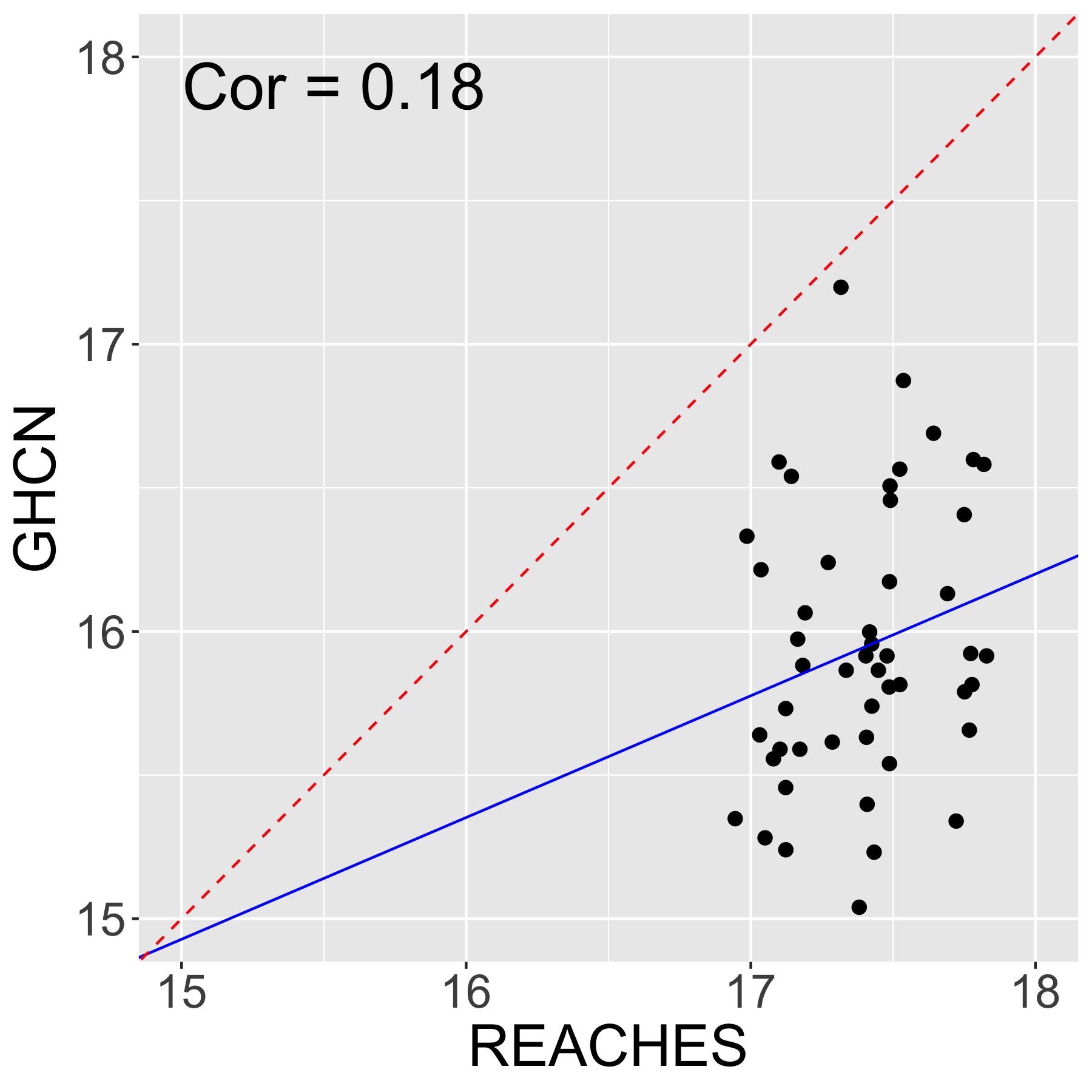}
        \caption{}
    \end{subfigure}
    \hfill
    \begin{subfigure}{0.32\linewidth}
        \includegraphics[width=\textwidth]{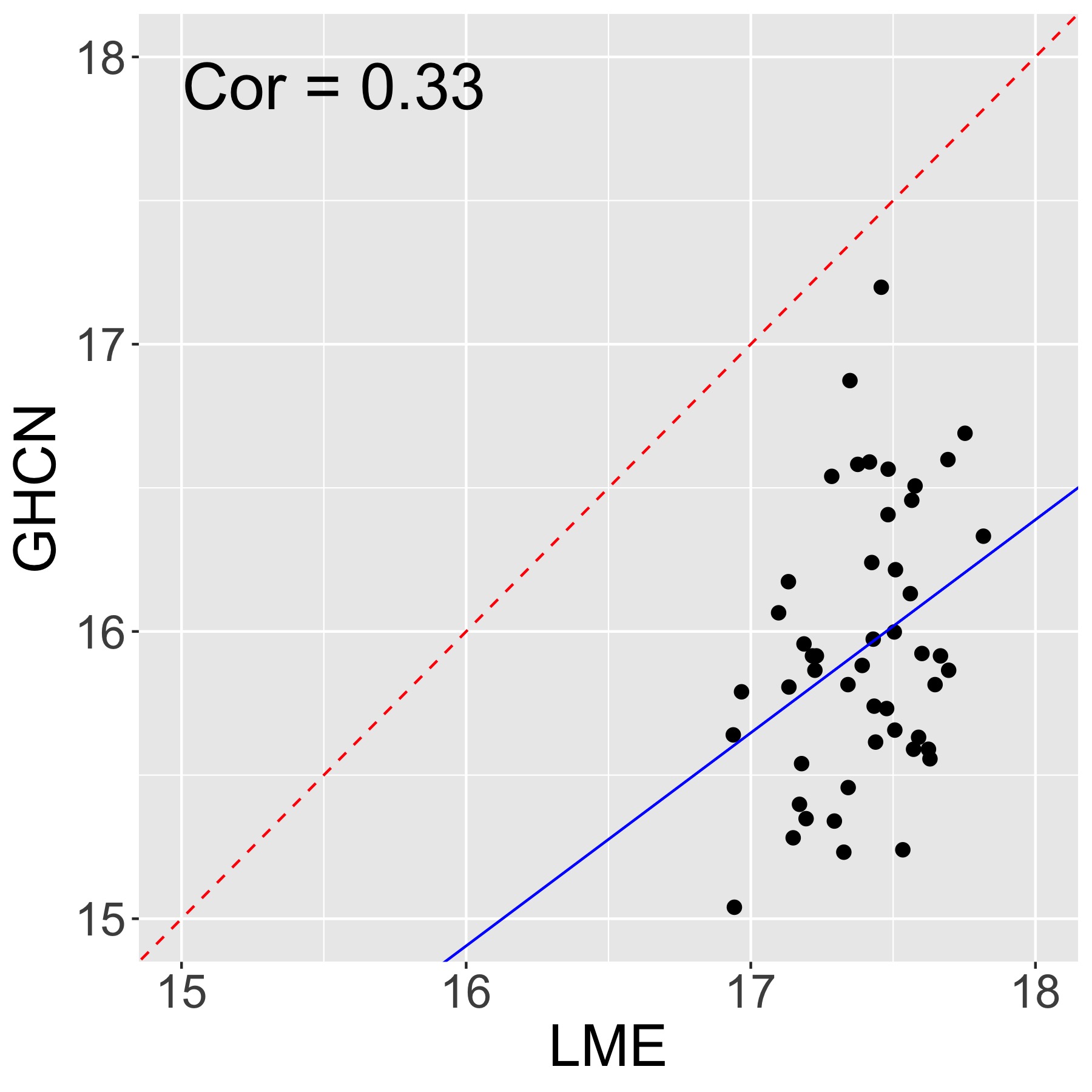}
        \caption{}
    \end{subfigure}
    \hfill
    \begin{subfigure}{0.32\linewidth}
        \includegraphics[width=\textwidth]{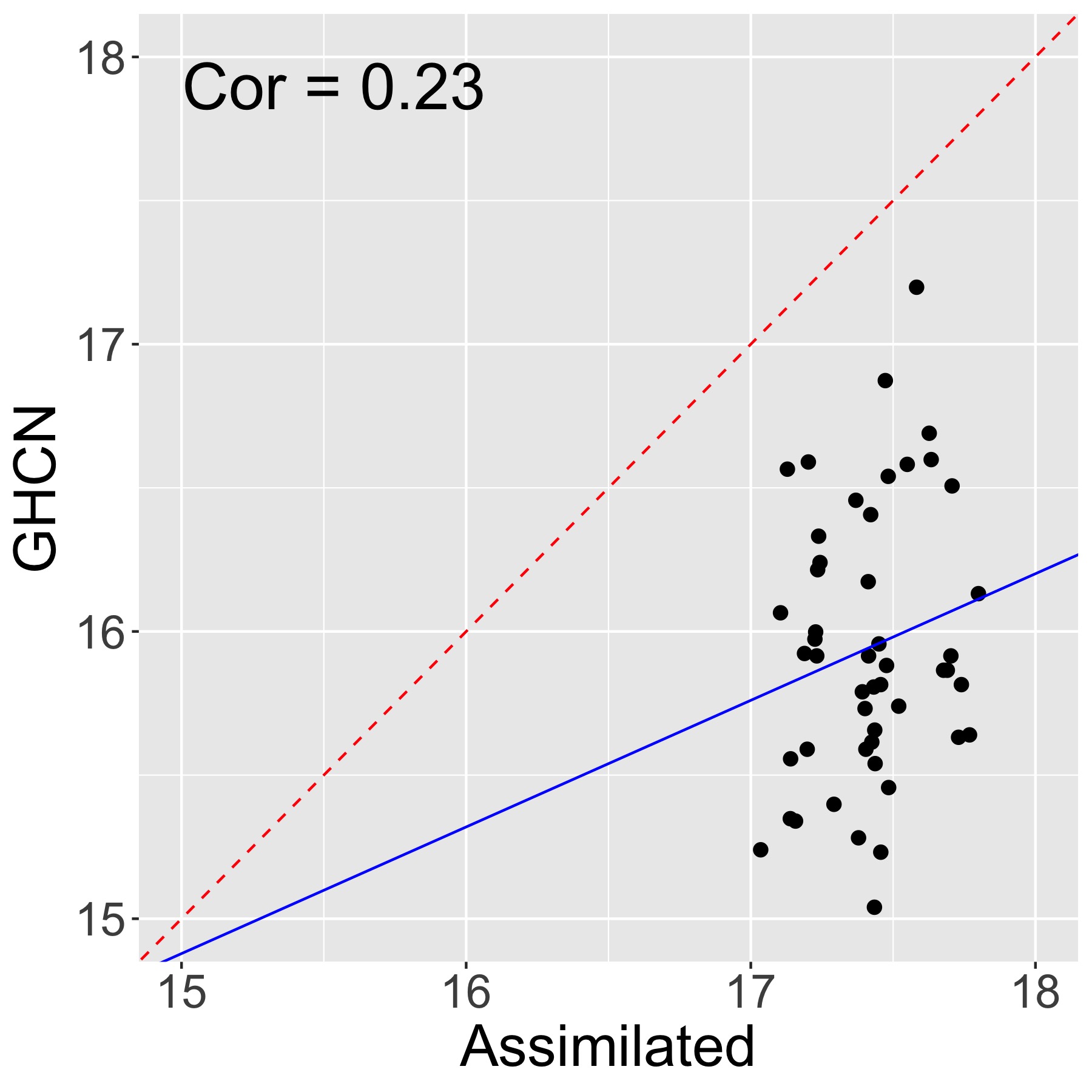}
        \caption{}
    \end{subfigure}
  \caption{Scatter plots between GHCN temperatures and estimates in Shanghai from 1861 to 1911: (a) GHCN vs.\ Celsius-scale REACHES; (b) GHCN vs.\ LME; (c) GHCN vs.\ Assimilated.}
    \label{ghcnn1}
\end{figure}

\begin{figure}[p]
    \centering
    \begin{subfigure}{0.32\linewidth}
        \includegraphics[width=\textwidth]{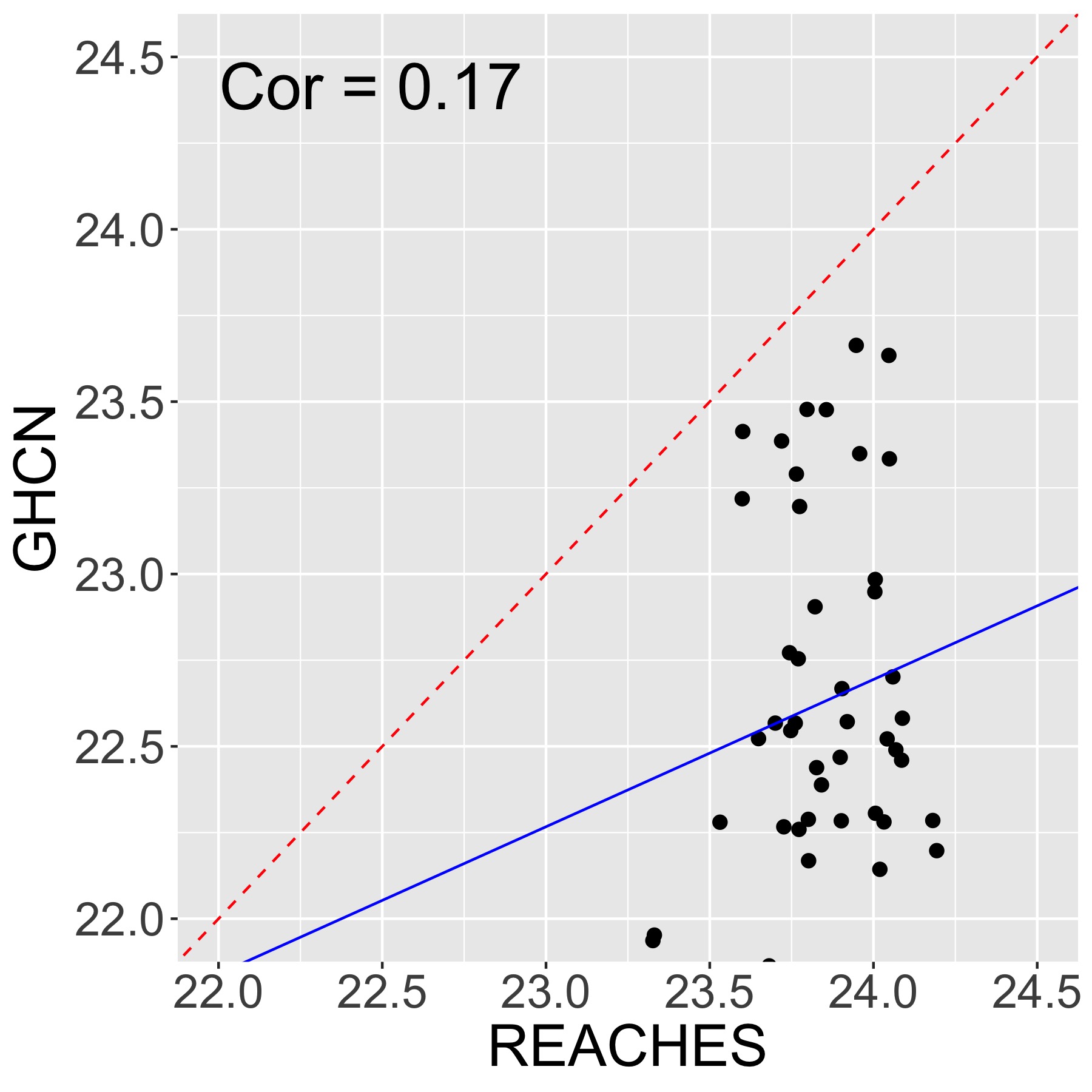}
        \caption{}
    \end{subfigure}
    \hfill
    \begin{subfigure}{0.32\linewidth}
        \includegraphics[width=\textwidth]{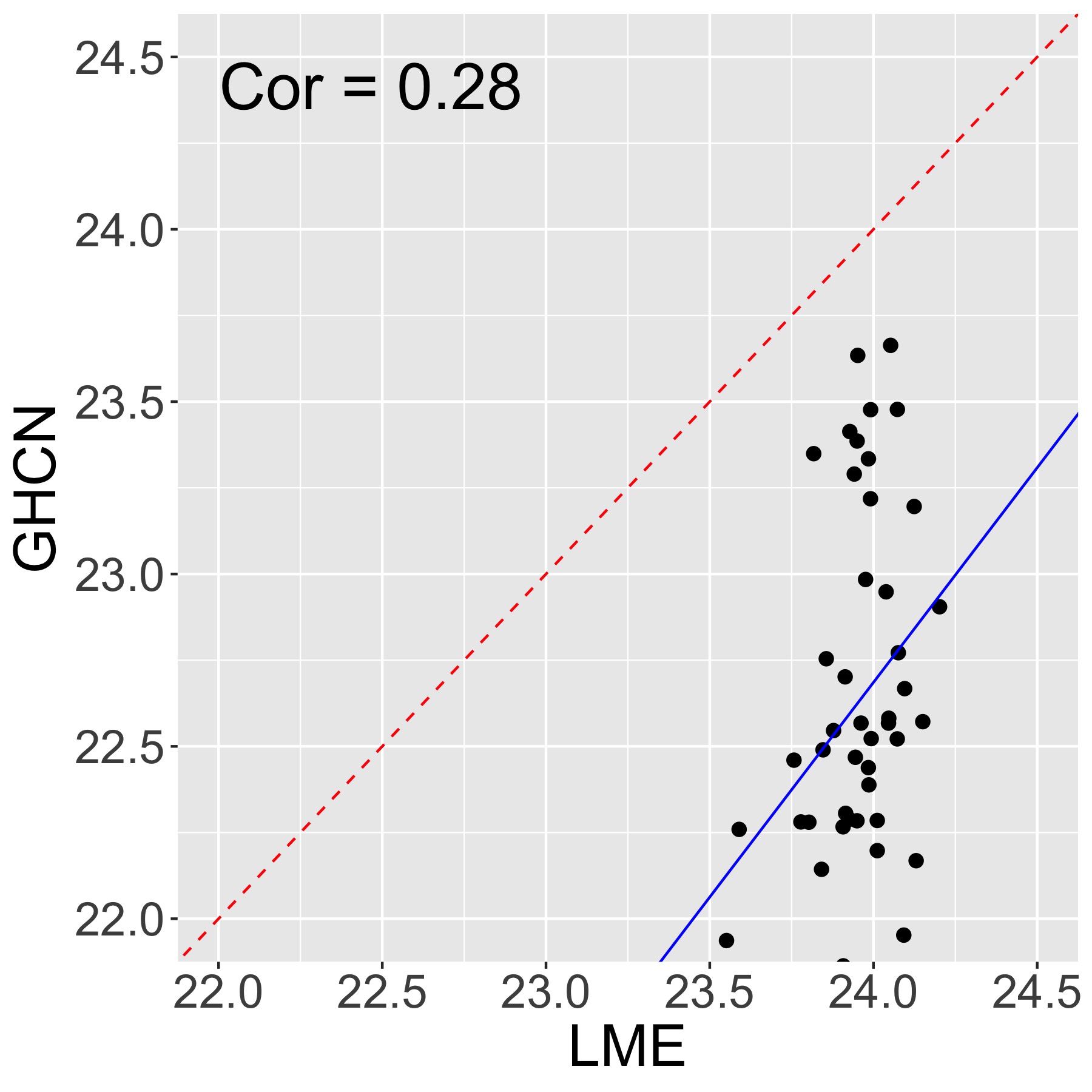}
        \caption{}
    \end{subfigure}
    \hfill
    \begin{subfigure}{0.32\linewidth}
        \includegraphics[width=\textwidth]{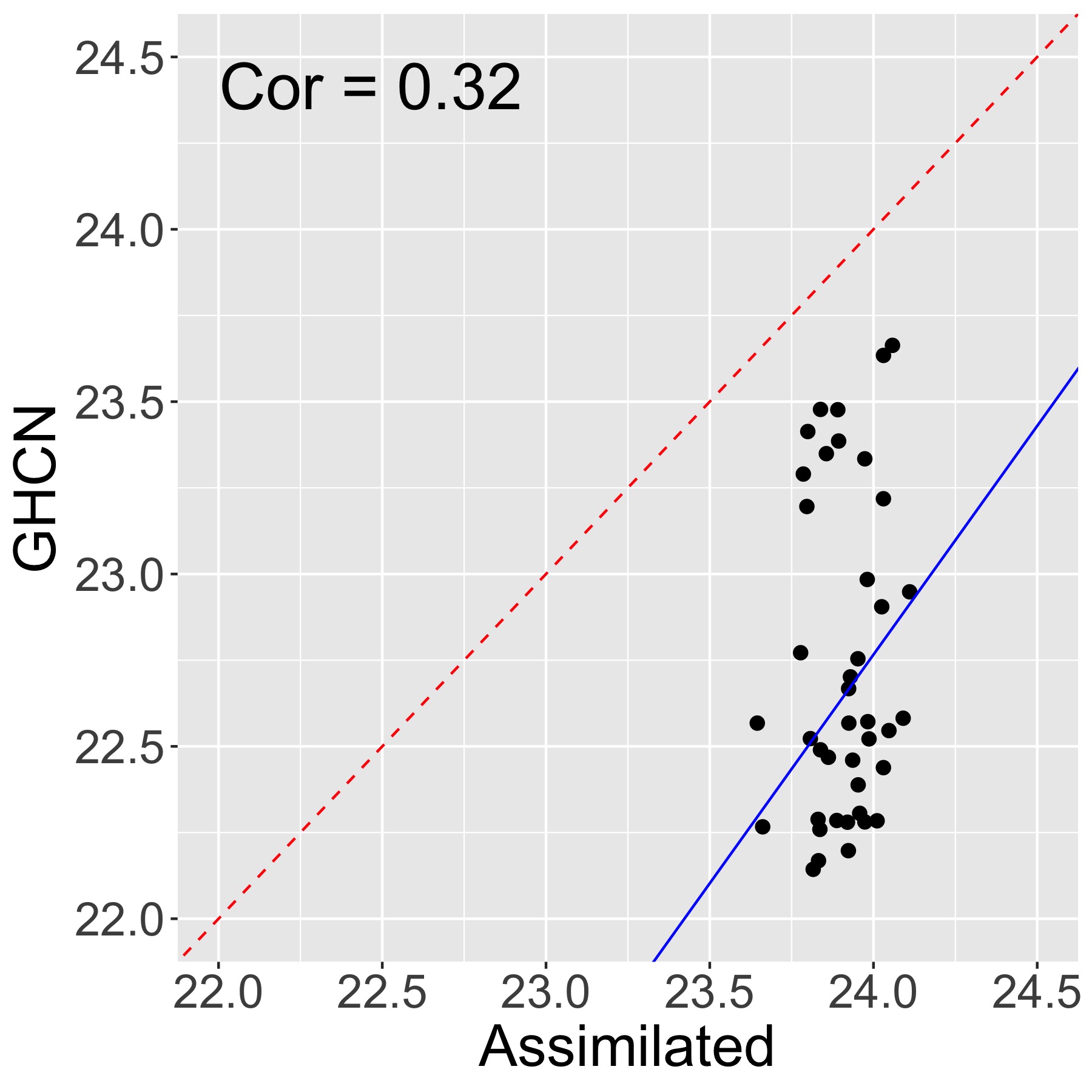}
        \caption{}
    \end{subfigure}
 \caption{Scatter plots between GHCN temperatures and estimates in Hong Kong from 1861 to 1911: (a) GHCN vs.\ Celsius-scale REACHES; (b) GHCN vs.\ LME; (c) GHCN vs.\ Assimilated.}
    \label{ghcnn2}
\end{figure}

To assess the reliability of our predictions, we convert the monthly GHCN data into annual aggregates and compare them against three methodologies: Celsius-transformed REACHES predictions generated through our kriging and quantile mapping method, LME simulation outputs, and Bayesian assimilated predictions. Figures~\ref{ghcnn}, \ref{ghcnn1}, and \ref{ghcnn2} depict scatter plots of GHCN data against the estimates from the three methods in Beijing, Shanghai, and Hong Kong, respectively. Generally, the temperature ranges produced by these three methods are narrower than those recorded in the GHCN, reflecting the inherent limitations of historical and model-based data. Except in Beijing, temperatures from all three methods tend to be higher than those from the GHCN.

While the Celsius-transformed REACHES predictions outperform the LME simulation outputs in Beijing in terms of correlation coefficients, they do not perform as well in Shanghai and Hong Kong. Our Bayesian assimilated predictions, however, further integrate information from the LME simulations, resulting in improved performance in Shanghai and Hong Kong. Specifically, compared to the Celsius-transformed REACHES predictions, our Bayesian assimilation method increases the correlation coefficient from 0.18 to 0.23 in Shanghai and 0.17 to 0.32 in Hong Kong. However, it decreases slightly from 0.50 to 0.45 in Beijing. Despite the slight decrease in Beijing, the correlation remains high, underscoring the robustness of our method. The overall improvement in correlation in Shanghai and Hong Kong confirms the effectiveness of integrating historical proxy data with climate model simulations, particularly in regions where individual methods perform less effectively.

The higher correlation in Beijing compared to Shanghai and Hong Kong may be attributed to several factors. One possibility is that Beijing's inland continental climate might exhibit temperature variations that are more pronounced and thus more accurately recorded in historical documents. In contrast, the coastal climates of Shanghai and Hong Kong may be influenced by local factors that are less consistently documented.

\section{Discussion and Conclusions}
\label{sec6}

In this article, we introduced a comprehensive Bayesian data assimilation framework to reconstruct historical temperatures in China during the Ming and Qing dynasties. By combining qualitative historical records from the REACHES dataset and climate model simulations from the LME, our method can produce spatially continuous temperature reconstructions at any point across East China, going beyond earlier studies like Wang et al. (2024), where reconstructions were confined to 15 subregions.

A central feature of our method is the geostatistical modeling of ordinal temperature indices as a zero-mean Gaussian process with interval censoring. This method accounts for data that are missing, not at random, by recognizing that the absence of a record often implies normal conditions. We applied quantile mapping to convert the estimated temperature indices into Celsius units. We then used a data assimilation procedure that combines a nonstationary time series prior, derived from LME simulations, with the Celsius-scale REACHES reconstruction through the Kalman filter and smoother algorithms. This process accounts for heterogeneous means and autocorrelations, substantially improving the accuracy of temperature estimates over the Ming and Qing dynasties.

Comparative analysis with GHCN instrumental records showed that our assimilated data generally outperformed standalone Celsius-scale REACHES and LME data. However, the enhancement beyond the LME data is modest in some locations, likely due to high local variability and limited GHCN observations.

Our framework also contributes valuable insights into historical climate variability, capturing significant climatic events such as the cooling period during the late Ming dynasty associated with the Little Ice Age. This agreement reinforces the reliability of our long-term reconstructions from qualitative documentary sources via REACHES data.



\bibliography{refs.bib}
\bibliographystyle{apalike}

\end{document}